\documentclass{amsart}
\usepackage{amsaddr}
\usepackage[T1]{fontenc}
\usepackage[utf8]{inputenc}

\usepackage[english]{babel}
\usepackage{amsmath}
\usepackage{amsthm}
\usepackage{amssymb}
\usepackage{graphicx}
\usepackage{url}
\usepackage{amsmath}
\usepackage{amssymb}
\usepackage{amsthm}
\usepackage{hyperref}
\usepackage{pgfplots}
\usepackage{subfig}
\usepackage{mathtools}

\DeclareMathOperator{\Diver}{Div}
\DeclareMathOperator{\Grad}{Grad}

\DeclareMathOperator{\trace}{tr}

\DeclareMathOperator{\diag}{diag}

\usepackage{mathtools}
\usepackage{pgfplots}
\pgfplotsset{/pgf/number format/use comma,compat=newest}

\newcommand{\R}{\mathbb{R}}

\newcommand{\vect}[1]{\boldsymbol{#1}}
\newcommand{\tens}[1]{\mathsf{#1}}

\begin{document}

\title[Shape transitions in a residually stressed sphere]{Shape transitions in a soft incompressible sphere with residual stresses}

\author{Davide Riccobelli, Pasquale Ciarletta}
\address{MOX -- Dipartimento di Matematica, Politecnico di Milano, Milano, Italy}
\email{davide.riccobelli@polimi.it, pasquale.ciarletta@polimi.it}


\begin{abstract}
Residual stresses may appear in elastic bodies due to the formation of misfits in the micro-structure, driven by plastic deformations, thermal or growth processes. They are especially widespread in living matter, resulting from the dynamic remodelling processes aiming at optimizing the overall structural response to environmental physical forces.  
From a mechanical viewpoint, residual stresses are classically modelled through the introduction of a virtual incompatible state that collects the local relaxed states around each material point. In this work, we instead employ an alternative approach based on a strain energy function that constitutively depends only on the deformation gradient and the residual stress tensor. In particular, our objective is to study the morphological stability of an incompressible sphere, made of a neo-Hookean material and subjected to given distributions of residual stresses. 
The boundary value elastic problem  is studied with analytic and numerical tools. Firstly, we perform  a linear stability analysis on the pre-stressed solid sphere using the method of incremental deformations. The marginal stability  conditions are given as a function of a control parameter, being the dimensionless variable that represents the characteristic intensity of the residual stresses. Secondly, we perform  finite element simulations using a mixed formulation in order to investigate  the post-buckling morphology in the fully nonlinear regime. 
Considering different initial distributions of the residual stresses, we find that different  morphological transitions occur around the material domain where the hoop residual stress reaches its maximum compressive value. The loss of spherical symmetry is found to be controlled by the mechanical and geometrical properties of the  sphere, as well as on the spatial distribution of the residual stress.
The results provide useful guidelines in order to design morphable soft spheres, for example by controlling the residual stresses through active deformations. They finally suggest a viable solution for the nondestructive characterization of residual stresses in soft tissues, such as solid tumors.

\end{abstract}


\maketitle

\section{Introduction}

Mechanical  stresses may be present inside elastic solid materials even in the absence of external forces:  they are commonly known as  residual stresses \cite{Hoger1985, hoger1986determination}. These stresses result from the presence of micro-structural misfits, for example after  plastic deformations (e.g. in metals),  thermal processes (e.g. quick solidification in glass) or  differential growth within biological tissues \cite{rodriguez1994stress}. Indeed, it is well acknowledged that there exists a mechanical feedback in many biological processes, e.g. in cell mitosis \cite{bao2003cell, montel2011stress}. Thus, living tissues can adapt their structural response to the external mechanical stimuli by generating residual stresses either in physiological conditions (e.g. within  arteries or the gastro-intestinal tract \cite{chuong1986residual, Dou2006, wang2014residual}) or pathological situations (e.g. solid tumors \cite{stylianopoulos2012causes, dolega2017cell, Ambrosi2017}). Moreover, residual  stresses can accumulate reaching  a critical threshold beyond which a morphological transition is triggered, possibly leading to complex pattern formation, such as  wrinkling, creasing or folding \cite{li2012mechanics,ciarletta2014pattern}.

Several studies about mechanical instabilities in soft materials have been carried out in the last decades. The stability of spherical elastic shells has been  studied with respect to the application of an external \cite{wesolowski1967stability,wang1972stability} or internal pressure \cite{Hill1976, haughton1978incremental}. More recently, the influence of residual stresses on stability in growing spherical shells \cite{amar2005growth} as well as in spherical solid tumor \cite{ciarletta2013buckling} has been addressed.

Residual stresses are classically modeled by performing a multiplicative decomposition of the deformation gradient \cite{rodriguez1994stress}. The key point of this method is the multiplicative decomposition of the deformation gradient $\tens{F}$ into two parts, being $\tens{F}=\tens{F}_e\tens{F}_o$, in which the tensor $\tens{F}_o$ defines  the natural state of the material free of  any geometrical constraint,  whereas $\tens{F}_e$ is the elastic deformation tensor  restoring the geometrical compatibility under the action of the external forces.

The main drawback of this method is the necessity of the  \textit{a priori} knowledge of the natural state,  since it is not often physically accessible. Indeed, from an experimental viewpoint, its determination would require several cuttings (ideally infinite) on the elastic body in order to release all the underlying residual stresses \cite{Dou2006, wang2014residual, stylianopoulos2012causes, Ambrosi2017}.

In this work, we employ an alternative approach based on a strain energy function that constitutively depends on both the deformation gradient and the residual stress tensor in the reference configuration \cite{shams2011initial}. In particular, our objective is to study the morphological stability of an incompressible sphere, naturally made of a neo-Hookean material and subjected to given distributions of residual stresses.

The work is organized as follows. Firstly,  we introduce the hyperelastic model for a pre-stressed material, defining the constitutive assumptions as a function of given distributions of residual stresses. Secondly, we apply the theory of incremental deformations in order to study the linear stability of a pre-stressed solid sphere with respect to the underlying residual stresses. Finally, we implement a numerical algorithm using the mixed finite element method in order to approximate the fully non-linear elastic solution. In the last section we discuss the results of the linear and non-linear analysis, together with some concluding remarks.

\section{The elastic model}

Let us consider a soft residually-stressed solid sphere composed of an incompressible hyperelastic material in a reference configuration $\Omega\subset\mathbb{E}^3$, where $\mathbb{E}^3$ is the three-dimensional Euclidean space. We use a spherical coordinate system in the reference configuration so that the material position vector is given by 
\[
\vect{X}=(R \sin(\Theta)\cos(\Phi),R \sin(\Theta)\sin(\Phi),\,R \cos(\Theta))
\]
where $R$ is the radial coordinate, $\Theta$ is the polar angle and $\Phi$ is the azimuthal angle.

We define the domain $\Omega$ as the set such that
\[
\Omega=\left\{\vect{X}\in\mathbb{E}^3\;|\;R<R_o\right\},
\]
so that $R_o$ is the radius of the solid sphere. We indicate with $\vect{e}_R,\,\vect{e}_\Theta$ and $\vect{e}_\Phi$ the local orthonormal vector basis.

\subsection{Constitutive assumptions}

Indicating with $\vect{x}=\vect{\varphi}(\vect{X})$ the spatial position vector, so that $\vect{\varphi}$ is the deformation field, we assume that the strain energy density of the body $\psi$ is a function depending on both the deformation gradient $\tens{F}=\Grad\vect{\varphi}$ and the Cauchy stress $\tens{\Sigma}$ in the reference configuration (i.e. the residual stress \cite{Hoger1985}): 
\begin{equation}
\label{eqn:strainenergy}
\psi=\psi(\tens{F},\,\tens{\Sigma}),
\end{equation}
as previously proposed in \cite{shams2011initial}.

Hence, the first Piola--Kirchhoff stress tensor $\tens{S}$ and the Cauchy stress tensor $\tens{T}$ are given by
\begin{equation}
\label{eqn:nominalecauchy}
\tens{S}(\tens{F},\,\tens{\Sigma})=\frac{\partial\psi}{\partial\tens{F}}\left(\tens{F},\,\tens{\Sigma}\right)-p\tens{F}^{-1},\qquad\tens{T}(\tens{F},\,\tens{\Sigma})=\tens{F}\tens{S}
\end{equation}
where $p$ is the Lagrangian multiplier that enforces the incompressibility constraint $\det\tens{F}=1$.

Hence, the fully non-linear problem in the quasi-static case reads
\begin{equation}
\label{eqn:fulnonlin}
\Diver \tens{S}=\vect{0}.
\end{equation}
where $\Diver$ denotes the divergence operator in material coordinates; the boundary conditions are
\begin{equation}
\label{eqn:fulnonlinBC}
\tens{S}^T\vect{e}_R=\vect{0} \qquad\text{when }R=R_o
\end{equation}
where $\vect{u}(\vect{X})=\vect{\varphi}(\vect{X})-\vect{X}$ is the displacement vector field.

When we evaluate the Piola--Kirchhoff stress in the reference configuration, we obtain the residual stress $\tens{\Sigma}$, i.e. setting $\tens{F}$ equal to the identity tensor $\tens{I}$ in Eq.~\eqref{eqn:nominalecauchy}, we get
\begin{equation}
\label{eqn:ISCC}
\tens{\Sigma}=\frac{\partial\psi}{\partial\tens{F}}\left(\tens{I},\,\tens{\Sigma}\right)-p_0\tens{I};
\end{equation}
this relation represents the \emph{initial stress compatibility condition} \cite{shams2011initial, gower2015initial, gower2016new}, where $p_0$ is a scalar field corresponding to the pressure field in the unloaded case.

Moreover, since $\tens{\Sigma}$ is the Cauchy stress tensor in the reference configuration, the balance of the linear and the angular momentum impose
\begin{equation}
\label{eqn:equilibriostressres}
\Diver \tens{\Sigma}=\vect{0},\qquad\tens{\Sigma}=\tens{\Sigma}^T\qquad \text{in }\Omega,
\end{equation}
together with the following boundary conditions
\begin{equation}
\label{eqn:stildebound}
\Sigma_{RR}=\Sigma_{\Theta R}=\Sigma_{\Phi R}=0\qquad\text{for }R=R_o.
\end{equation}

From Eqs.~\eqref{eqn:equilibriostressres}-\eqref{eqn:stildebound}, it is possible to prove that \cite{hoger1986determination}
\[
\int_{\Omega}\tens{\Sigma}\,d\mathcal{L}^{3}(\vect{X})=\tens{0},
\]
so that the residual stress field must be inhomogeneous, with zero mean value.

We also assume that the strain energy density depends on the choice of the reference configuration only through the functional dependence on ${\tens{\Sigma}}$. Thus, we impose the  \emph{initial stress reference independence} (see \cite{gower2015initial, gower2016new} for further details), reading 
\begin{equation}
\label{eqn:ISRI}
\psi\left(\tens{F}_1\tens{F}_2,\,\tens{\Sigma}\right)=\psi\left(\tens{F}_1,\,\tens{T}\left(\tens{F}_2,\,\tens{\Sigma}\right)\right).
\end{equation}

The Eq. \eqref{eqn:ISRI} must hold for all second order tensor $\tens{F}_1$, $\tens{F}_2$ with positive determinant and for all the symmetric tensors $\tens{\Sigma}$.

The general material with a strain energy given by Eq.~\eqref{eqn:strainenergy}, such that the material behaviour is isotropic in absence of residual stress, i.e. for $\tens{\Sigma}=\tens{0}$, may depend up to ten independent invariants \cite{shams2011initial}.

A simple possible choice for the strain energy density which satisfies both the initial stress compatibility condition and the initial stress reference independence is the one corresponding to an \emph{initially stressed neo--Hookean material}. The strain energy of such material is constructed so that if a virtual relaxed state exists \cite{johnson1995use}, then it naturally behaves as a neo--Hookean material with a given shear modulus $\mu$. In the following we briefly sketch how this strain energy is obtained (see \cite{gower2015initial} for a detailed derivation).

Let us introduce the following five invariants:
\begin{gather*}
I_1 = \trace \tens{C}, \quad
J_1 = \trace \left(\tens{\Sigma} \tens{C}\right),\quad I_{\tens{\Sigma}1}=\trace\tens{\Sigma},\quad
I_{\tens{\Sigma}2}=\frac{(\trace\tens{\Sigma})^2-
\trace\tens{\Sigma}^2}{2},\quad
I_{\tens{\Sigma}3}=\det\tens{\Sigma},
\end{gather*}
where $\tens{C} = \tens{F}^T\tens{F}$ is the right Cauchy--Green tensor.

Assuming that the material behaves as an incompressible neo-Hookean body, its strain energy density is given by
\begin{equation}
\label{eqn:neohookprestress}
\psi(\mathsf{F},\,\tens{\Sigma})=\frac{\mu}{2}(\trace(\widetilde{\mathsf{B}}\mathsf{C})-3),
\end{equation}
here $\mathsf{C}$ is the right Cauchy-Green strain tensor, $\widetilde{\mathsf{F}}$ is the deformation gradient from the virtual unstressed state to the reference configuration, $\widetilde{\mathsf{B}}=\widetilde{\mathsf{F}}\widetilde{\mathsf{F}}^{\mathsf{T}}$ and $\mu$ is the shear modulus of the material in absence of residual stresses.

So, substituting Eq.~\eqref{eqn:neohookprestress} in Eq.~\eqref{eqn:nominalecauchy}, the initial stress $\tens{\Sigma}$ is given by
\begin{equation}
\label{eqn:stressresiduoneohook}
\tens{\Sigma}=\mu\widetilde{\mathsf{B}}-\widetilde{p}\mathsf{I}.
\end{equation}

Imposing the incompressibility constraint on the deformation gradient $\widetilde{\tens{F}}$, we get $\det(\mu\widetilde{\mathsf{B}})=\mu^3=\det(\tens{\Sigma}+\widetilde{p}I)$. Thus, $\widetilde{p}$ is the real root of the following polynomial:
\begin{equation}
\label{eqn:equazioneptilde}
\tilde{p}^3+\tilde{p}^2I_{\tens{\Sigma}1}+\tilde{p}I_{\tens{\Sigma}2}+I_{\tens{\Sigma}3}-\mu^3=0.
\end{equation}

Hence, multiplying Eq.~\eqref{eqn:stressresiduoneohook} by $\mathsf{C}$ on the right and taking the trace on both sides, we obtain
\begin{equation}
\label{eqn:J1}
\trace(\widetilde{\mathsf{S}}\mathsf{C})=\mu\trace(\widetilde{\mathsf{B}}\mathsf{C})-
\widetilde{p}I_{1}.
\end{equation}

Substituting Eq.~\eqref{eqn:J1} in Eq.~\eqref{eqn:neohookprestress}, we obtain the strain energy of an initially stressed Neo--Hookean body:
\begin{equation}
\label{eqn:strainenergyexpr}
\psi\left(I_1,\,J_1,\,I_{\tens{\Sigma}1},\,I_{\tens{\Sigma}2},\,I_{\tens{\Sigma}3}\right)=\frac{1}{2}(J_1+\widetilde{p}I_1-3\mu).
\end{equation}
where $\tilde{p}$ is the only real root of Eq. \eqref{eqn:equazioneptilde}. It is given by \cite{gower2015initial}
\[
\tilde{p}=\frac{1}{3}\left[T_3+\frac{T_1}{T_3}-I_{\tens{\Sigma}1}\right],
\]
where
\begin{gather*}
T_1=I_{\tens{\Sigma}1}^2-3 I_{\tens{\Sigma}2},\\
T_2=I_{\tens{\Sigma}1}^3-\frac{9}{2} I_{\tens{\Sigma}1} I_{\tens{\Sigma}2}+\frac{27}{2} \left(I_{\tens{\Sigma}3}-\mu ^3\right),\\
T_3=\sqrt[3]{\sqrt{T_2^2-T_1^3}-T_2}.
\end{gather*}

In this setting, it is possible to prove that the pressure field in the reference configuration is given by $p=\tilde{p}$ \cite{gower2015initial}.

In the following, we use symmetry arguments to discuss a few possible choices for the distribution of the residual stresses.

\subsection{Residual stress distribution}

We assume that the residual stress $\tens{\Sigma}$ depends only on the variable $R$. Hence the system of equations given by Eq.~\eqref{eqn:equilibriostressres} reduces to
%
%
\begin{equation}
\label{eqn:stildeeq}
\left\{
\begin{aligned}
&\frac{\partial \Sigma_{RR}}{\partial R}+\frac{2}{R}(\Sigma_{RR}-\Sigma_{\Theta\Theta})=0,\\
&\Sigma_{R\Theta}=\Sigma_{R\Phi}=\Sigma_{\Theta\Phi}=0;
\end{aligned}
\right.
.
\end{equation}

\begin{figure}[t!]
\begin{center}
\subfloat{
\includegraphics[width=0.45\textwidth]{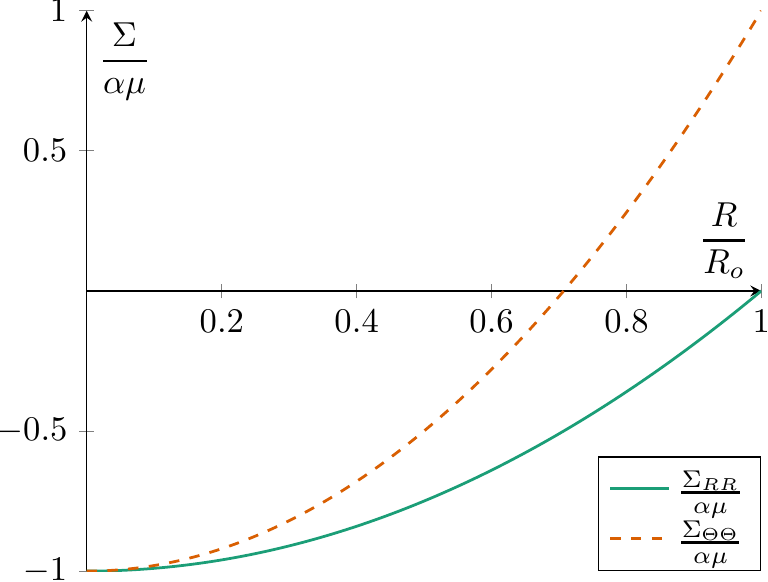}}\qquad
\subfloat{
\includegraphics[width=0.45\textwidth]{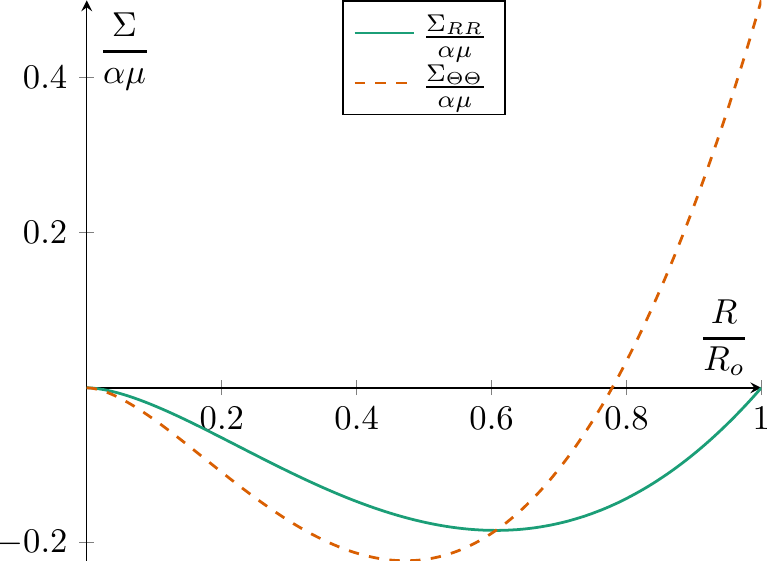}}
\caption{Plot of the radial (solid line) and hoop (dashed line) residual stress components normalized with respect to $\alpha \ \mu$ when $f(R)$ is given by Eq.~\eqref{eqn:fpolinomiale} (left) $f(R)$ is given by Eq.~\eqref{eqn:flog} (right). Both the dimensionless parameters $\beta$ and $\gamma$ are set equal to $2$.}
\label{fig:res_stress}
\end{center}
\end{figure}

Then, being $f(R)$ the radial component of the residual stress, the tensor $\tens{\Sigma}$ is given by
\[
\tens{\Sigma}=
\diag\left(
f(R),\,\frac{R}{2}f'(R)+f(R),\,\frac{R}{2}f'(R)+f(R),
\right)
\]
where $f:[0,R_o]\rightarrow\R$ is such that $f(R_o)=0$ in order to satisfy automatically Eq.~\eqref{eqn:stildeeq}.

In the following, we will focus on two possible choices for the function $f$:
\begin{gather}
\label{eqn:fpolinomiale}
\text{\textbf{case (a)}}:f(R)=\alpha\mu\frac{R^\beta-R_o^\beta}{R_o^\beta},\\
\label{eqn:flog}
\text{\textbf{case (b)}}:f(R)=\alpha\mu\left(\dfrac{R}{R_o}\right)^\gamma\log\left(\dfrac{R}{R_o}\right),
\end{gather}
where $\alpha,\,\beta$ and $\gamma$ are real dimensionless parameters with $\beta,\,\gamma>1$. The corresponding residual stress components are depicted in Fig.~\ref{fig:res_stress}.

In the next section we apply the theory of incremental deformations in order to study the stability of the residually stressed configuration with respect to the magnitude of the underlying residual stresses expressed by the dimensionless parameters $\alpha,\,\beta$ and $\gamma$.

\section{Incremental problem and linear stability analysis}

\subsection{Structure of the incremental equations}
In order to study the linear stability of the undeformed configuration with respect to the intensity of the residual stresses, we use the method of the incremental elastic deformations~\cite{ogden1997non}. We denote with $\delta\vect{u}$ the incremental displacement vector and with $\tens{\Gamma}$ the gradient of the vector field $\delta\vect{u}$, namely $\tens{\Gamma}=\Grad\delta\vect{u}$.

The linearized incremental Piola--Kirchhoff stress tensor reads
\begin{equation}
\label{eqn:IncrNomStress}
\delta\tens{S}=\mathcal{A}_0^1:\tens{\Gamma}+p\tens{\Gamma}-q\tens{I}
\end{equation}
where $q$ is the increment of the Lagrangian multiplier $p$ and
\[
\left(\mathcal{A}_0^1:\tens{\Gamma}\right)_{ij}\coloneqq A_{0ijhk}^1\Gamma_{kh}=\left.\frac{\partial\psi}{\partial F_{ji}\partial F_{kh}}\right|_{\tens{F}=\tens{I}}\Gamma_{kh},
\]
with $\mathcal{A}_0$ being the fourth order tensor of the elastic moduli, and summation over repeated subscripts is assumed.

From Eq.~\eqref{eqn:strainenergyexpr} and following \cite{shams2011initial}, we get
\[
A_{0ijhk}^1=\delta_{jk}(2\psi,_{I_1}\delta_{ih}+\tens{\Sigma}_{ih}),
\]
where $\delta_{ij}$ is the Kronecker delta the comma denotes the partial derivative.

Hence, the incremental equilibrium equation is given by
\begin{equation}
\label{eqn:IncrEquiEq}
\Diver \delta\tens{S}=\vect{0},
\end{equation}
and the boundary conditions read
\begin{equation}
\label{eqn:ContornoDeltaS}
\delta\tens{S^T}\vect{e}_R=\vect{0}\qquad\text{at }R=R_o.
\end{equation}

The incompressibility of the incremental deformation is given by the constraint
\begin{equation}
\label{eqn:IncompDeltaS}
\trace\tens{\Gamma}=0.
\end{equation}

We assume an axis-symmetric incremental displacement vector given by
\[
\delta\vect{u}=u(R,\,\Theta)\vect{e}_{R}+v(R,\,\Theta)\vect{e}_{\Theta}.
\]
This choice is motivated by the fact that, imposing a general incremental displacement vector, the resulting governing equations in the azimuthal direction decouple \cite{wesolowski1967stability, wang1972stability}, thus not influencing the linearized bifurcation analysis.

Hence, the incremental displacement gradient is given by
\[
\tens{\Gamma}=
\begin{bmatrix}
 u,_{R} & \dfrac{u,_{\Theta}-v}{R} & 0 \\
 v,_{R} & \dfrac{u+v,_{\Theta}}{R} & 0 \\
 0 & 0 & \dfrac{u+\cot (\Theta) v}{R} \\
\end{bmatrix}.
\]

In order to build a robust numerical procedure to solve the incremental boundary value problem, we first rewrite  Eqs.~\eqref{eqn:IncrEquiEq}-\eqref{eqn:IncompDeltaS} using a more convenient form, known as Stroh formulation.

\subsection{Stroh formulation}
Since the residually stressed material is inhomogeneous only in the radial direction, we study the bifurcation problem by assuming variable separation for the incremental fields \cite{norris2012elastodynamics}, namely
\begin{gather}
\label{eqn:uPm}
u(R,\,\Theta)=U(R)P_m(\cos\Theta),\\
\label{eqn:vPm}
v(R,\,\Theta)=V(R)\frac{1}{\sqrt{m(m+1)}}\frac{d P_m(\cos\Theta)}{d\Theta},\\
\label{eqn:SRRdeco}
\delta S_{RR}(R,\,\Theta)=s_{RR}(R)P_m(\cos\Theta),\\
\delta S_{R\Theta}(R,\,\Theta)=s_{R\Theta}(R)\frac{1}{\sqrt{m(m+1)}}\frac{d P_m(\cos\Theta)}{d\Theta},
\end{gather}
where $P_m(\Theta)$ denotes the Legendre polynomial of order $m$.

In order to write the incremental boundary value problem Eqs.~\eqref{eqn:IncrEquiEq}-\eqref{eqn:IncompDeltaS} in the Stroh formulation,  we introduce the displacement-traction vector $\vect{\eta}$, defined as
\[
\vect{\eta}(R)=\begin{bmatrix}
\vect{U}(R)\\
R^2\vect{T}(R)
\end{bmatrix},
\qquad\text{where}\quad
\vect{U}(R)=\begin{bmatrix}
U(R)\\
V(R)
\end{bmatrix},
\quad
\vect{T}(R)=\begin{bmatrix}
s_{RR}(R)\\
s_{R\Theta}(R)
\end{bmatrix}.
\]

An expression for $q$ is found by substituting Eq.~\eqref{eqn:IncrNomStress} in Eq.~\eqref{eqn:SRRdeco}, so that
\begin{equation}
\label{eqn:q}
\begin{gathered}
q=P_m(\cos (\Theta)) \left(U'(R) \left(2 \psi,_{I_1}+f(R)+p\right)-\delta S_{RR}(R)\right).
\end{gathered}
\end{equation}

Thus, using a well established procedure \cite{stroh1962steady}, we can use the definition of the linearized incremental Piola--Kirchhoff given by Eq.~\eqref{eqn:IncrNomStress}, the incremental equilibrium equations given by Eq.~\eqref{eqn:IncrEquiEq} and the linearized incompressibility constraint Eq.~\eqref{eqn:IncompDeltaS}  to obtain a first order system of ordinary differential equations, namely
\begin{equation}
\label{eqn:StrohEq}
\frac{d\vect{\eta}}{dR}=\frac{1}{R^2}\tens{N}\vect{\eta},
\end{equation}
where $\tens{N}(R)$ is the \emph{Stroh matrix} which has the following structure
\[
\tens{N}=\begin{pmatrix}
\tens{N}_1 &\tens{N}_2\\
\tens{N}_3 &-\tens{N}^{\tens{T}}_1
\end{pmatrix},
\]
where the sub-blocks read:
\begin{gather*}
\mathsf{N}_1=\begin{pmatrix}
 -2 R & \sqrt{m (m+1)} R \\
 -\frac{\sqrt{m (m+1)} p R}{f(R)+2 \psi,_{I_1}} & \frac{p R}{f(R)+2 \psi,_{I_1}}
\end{pmatrix},\\
\mathsf{N}_2=\begin{pmatrix}
 0 & 0 \\
 0 & \frac{1}{f(R)+2 \psi,_{I_1}}
\end{pmatrix},\quad\mathsf{N}_3=\begin{pmatrix}
 \nu_{1} &\nu_{2}\\
 \nu_{2} &\nu_{3}
\end{pmatrix}.
\end{gather*}

The expressions for the coefficients $\nu_1,\,\nu_2$ and $\nu_3$ are given by:
\begin{footnotesize}
\begin{gather*}
\nu_{1}=\frac{R^2((2 \psi,_{I_1}+f(R)) (4 (m^2+m+6) \psi,_{I_1}+(m^2+m+2) R
   f'(R)}{2 (2 \psi,_{I_1}+f(R))}+\\
+\frac{2 (m^2+m+6) f(R)+12 p)-2 m (m+1) p^2)}{2 (2 \psi,_{I_1}+f(R))},\\
\nu_2=\frac{R^2\sqrt{m (m+1)} \left(p^2-\left(2 \psi,_{I_1}+f(R)\right) \left(8 \psi,_{I_1}+R f'(R)+4 f(R)+3
   p\right)\right)}{2 \psi,_{I_1}+f(R)},\\
\nu_3=\frac{R^2\left(2 \psi,_{I_1}+f(R)\right) \left(m (m+1) \left(8 \psi,_{I_1}+R f'(R)+4 f(R)\right)+2 (2 m
   (m+1)-1) p\right)}{2 \left(2 \psi,_{I_1}+f(R)\right)}+\\
   -\frac{2 R^2p^2}{2 \left(2 \psi,_{I_1}+f(R)\right)}.
\end{gather*}
\end{footnotesize}

In the next section, we solve the Eq.~\eqref{eqn:StrohEq} by using the impedance matrix method.

\subsection{Impedance matrix method}

Let us briefly sketch the main theoretical aspects of this method \cite{biryukov1985impedance,biryukov1995surface}. We define a linear functional relation between $\vect{U}$ and $\vect{T}$, namely
\begin{equation}
\label{SZU}
R^2\vect{T}=\tens{Z}\vect{U}.
\end{equation}
where $\tens{Z}$ is the so called \emph{surface impedance matrix}.

By substituting Eq.~\eqref{SZU} in Eq.~\eqref{eqn:StrohEq}, we obtain
\begin{gather}
\label{eqn:U'}
\frac{d\vect{U}}{d R}=\frac{1}{R^2}(\tens{N}_1\vect{U}+\tens{N}_2\tens{Z}\vect{U}),\\
\label{eqn:ZU'}
\frac{d\tens{Z}}{dR}\vect{U}+\tens{Z}\frac{d\vect{U}}{dR}=\frac{1}{R^2}(\tens{N}_3\vect{U}+\tens{N}_4\tens{Z}\vect{U}).
\end{gather}
Thus, by substituting Eq.~\eqref{eqn:U'} in Eq.~\eqref{eqn:ZU'},  a Riccati differential  equation is found for $\tens{Z}$, being
\begin{equation}
\label{eqn:Riccati}
\frac{d\tens{Z}}{dR}=\frac{1}{R^2}\left(\tens{N}_3-\tens{N}^\tens{T}_1\tens{Z}-\tens{ZN}_1- \tens{ZN}_2\tens{Z}\right).
\end{equation}

 Let now us define $\tens{M}$ as the solution to the following problem
\begin{equation}
\label{eqn:conditional matrix}
\left\{
\begin{aligned}
&\frac{d}{dR}\tens{M}(R,\,R_o)-\frac{\tens{N}}{R^2}\tens{M}(R,\,R_o)=\tens{0}\\
&\tens{M}(R_o,\,R_o)=\tens{I}.
\end{aligned}
\right.
\end{equation}
where the matricant $\tens{M}(R,\,R_o)$ is a $4\times4$ matrix, called the \emph{conditional matrix}.

Since $\tens{M}$ is the solution of the problem given in Eq.~\eqref{eqn:conditional matrix}, from Eq.~\eqref{eqn:StrohEq} it is straightforward to show that
\begin{equation}
\label{eqn:EtaM}
\vect{\eta}(R)=\tens{M}(R,\,R_o)\vect{\eta}(R_o).
\end{equation}

Let us split the conditional matrix into four blocks as
\begin{equation}
\label{eqn:Mblock}
\tens{M}=\begin{bmatrix}
\tens{M}_1(R,\,R_o) &\tens{M}_2(R,\,R_o)\\
\tens{M}_3(R,\,R_o) &\tens{M}_4(R,\,R_o)\\
\end{bmatrix}.
\end{equation}

We can use two possible ways to construct the surface impedance matrix, either the \emph{conditional impedance matrix} $\tens{Z}^\text{c}(R,\,R_o)$ or the \emph{solid impedance matrix} $\tens{Z}^\text{s}(R)$ \cite{norris2010wave}.

In fact, considering that $\vect{T}(R_o)=\vect{0}$ and by using the Eqs.~\eqref{eqn:EtaM}-\eqref{eqn:Mblock}, we can define the conditional impedance matrix as $\tens{Z}^\text{c}(R,\,R_o)\coloneqq\tens{M}_3(R,\,R_o)\tens{M}_1^{-1}(R,\,R_o)$. Such a matrix is called conditional since it depends explicitly on its value at $R=R_o$. 

Conversely, the solid impedance matrix does not depend explicitly on its value at one point, but instead it ensures that the surface impedance matrix is well posed at the origin. 

Following \cite{norris2010wave}, we consider a Taylor series expansion of the solid impedance matrix $\tens{Z}^\text{s}(R)$ around $R=0$, namely 
\begin{equation}
\label{eqn:Ztaylor}
\tens{Z}^\text{s}(R)=\tens{Z}_0 + \tens{Z}_1 R+o(R),
\end{equation}
where $\tens{Z}_0$ is called \emph{central impedance matrix}.

From the Eq.~\eqref{eqn:Riccati}, the solid impedance matrix is well posed at the origin only if the central impedance matrix satisfies the following algebraic Riccati equation:
\[
\tens{N}_3(0)-\tens{N}^\tens{T}_1(0)\tens{Z}_0-\tens{Z}_0\tens{N}_1(0)- \tens{Z}_0\tens{N}_2(0)\tens{Z}_0=\tens{0};
\]
whose general solution is given by
\begin{equation}
\label{eqn:centralimp}
\tens{Z}_0=\delta \vect{e}_1\otimes\vect{e}_1,\qquad \delta\in\mathbb{R}.
\end{equation}

By substituting Eq.~\eqref{eqn:Ztaylor} in Eq.~\eqref{eqn:Riccati} and setting $R=R_c\ll 1$, we obtain the following algebraic Riccati equation
\begin{equation}
\label{eqn:Z1}
\begin{gathered}
\tens{0}=\tens{N}_3(R_c)-\tens{N}^{\tens{T}}_1(R_c)\tens{Z}_0-\tens{Z}_0 \tens{N}_1(R_c)
-\tens{Z}_0\tens{N}_2(R_c)\tens{Z}_0-R_c^2\tens{Z}_1\tens{N}_2(R_c)\tens{Z}_1+\\
-R_c\tens{Z}_1\left(\tens{N}_1(R_c) +\tens{N_2}(R_c)\tens{Z}_0+\frac{R_c}{2}\tens{I}\right)-R_c\left(\tens{N}^{\tens{T}}_1(R_c) +\tens{Z}_0\tens{N_2}(R_c)+\frac{R_c}{2}\tens{I}\right)\tens{Z}_1
\end{gathered}
\end{equation}
whose stable solution is the only one such that the eigenvalues of
\[
-R_c\left(\tens{N}_1(R_c) +\tens{N_2}(R_c)\tens{Z}_0+\frac{R_c}{2}\tens{I}\right)-R_c^2\tens{N}_2(R_c)\tens{Z}_1
\]
are all negative \cite{ciarletta2014torsion}.

In summary, the surface impedance method allows us to avoid the direct resolution of the boundary value problem given by Eqs.~\eqref{eqn:IncrEquiEq}-\eqref{eqn:IncompDeltaS} by using  a numerical integration of the Riccati equation given by Eq.~\eqref{eqn:Riccati}.

\subsection{Numerical procedure and results of the linear stability analysis}

The aim of this section is to implement a robust numerical procedure to analyze the onset of a morphological transition as a function of the dimensionless parameters $\alpha,\,\beta$ and $\gamma$ representing the magnitude and the spatial distribution of the residual stresses.

The solution of the incremental boundary value problem can be obtained by a numerical integration of the differential Riccati equation~\eqref{eqn:Riccati} using two different procedures.

First, the differential Riccati equation in  Eq.~\eqref{eqn:Riccati} can be integrated from $R_c$ to $R_o$ with starting value
\begin{equation}
\label{eqn:Zinizint}
\tens{Z}^\text{s}(R_c)=\tens{Z}_0+R_c\tens{Z}_1,
\end{equation}
given by the solid impedance matrix in Eq.~\eqref{eqn:Ztaylor}.

Using Eq.~\eqref{eqn:Zinizint}, we numerically solve  Eq.~\eqref{eqn:Riccati} by iterating on   the value $\alpha$ in Eqs.~\eqref{eqn:fpolinomiale}-\eqref{eqn:flog}, starting from $0$ until the stop condition
\begin{equation}
\label{eqn:stopconditionint}
\det\tens{Z}^\text{s}(R_o)=0,
\end{equation}
is reached, namely when the impedance matrix is singular and the incremental Eqs.~\eqref{eqn:IncrEquiEq} and \eqref{eqn:IncompDeltaS} admit a non-null solution that satisfies Eq.~\eqref{eqn:ContornoDeltaS}.

A second approach consists in integrating Eq.~\eqref{eqn:Riccati} by using the conditional impedance matrix $\tens{Z}^\text{c}(R,\,R_o)$. Since from Eq.~\eqref{eqn:EtaM} it can be shown that  $\tens{M}(R_o,\,R_o)=\tens{I}$, the definition of the conditional impedance matrix given by Eq.~\eqref{eqn:conditional matrix} allows us to set the following initial condition:
\begin{equation}
\label{eqn:Zinizext}
\tens{Z}^\text{c}(R_o,\,R_o)=\tens{0}.
\end{equation}

Analogously, we iteratively integrate Eq.~\eqref{eqn:Riccati} until the stop condition
\begin{equation}
\label{eqn:stopconditionext}
\det(\tens{Z}^\text{c}(R_c,\,R_o)-\tens{Z}_0-\tens{Z}_1 R_c)=0
\end{equation}
is reached. This condition corresponds to the existence of non-null solutions for the variable $\vect{U}$ by imposing the continuity of the incremental stress vector $\vect{T}$ at $R=R_c$. 

In both cases, in order to find the incremental displacement field, we perform a further integration of Eq.~\eqref{eqn:U'} using the procedure described in \cite{destrade2009bending}.

The two numerical schemes were implemented by using the software \emph{Mathematica 11.0} (Wolfram Research, Champaign, IL, USA) in order to identify the marginal stability curves as function of the dimensionless parameters $\alpha,\,\beta$ and $\gamma$.

\subsubsection{Case (a): exponential polynomial case}

\begin{figure}[t!]
\centering
\subfloat{\includegraphics[scale=0.85]{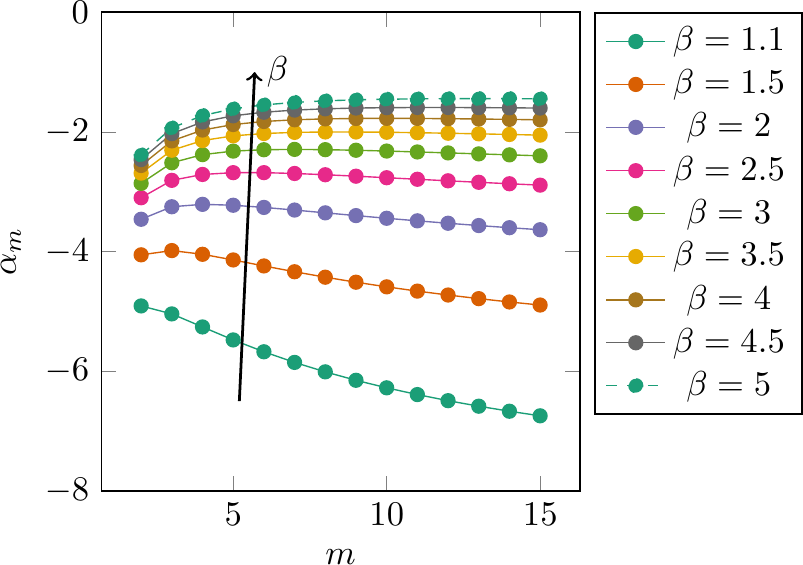}
			\label{fig:Ca1}
			}
		\subfloat{
			\includegraphics[scale=0.85]{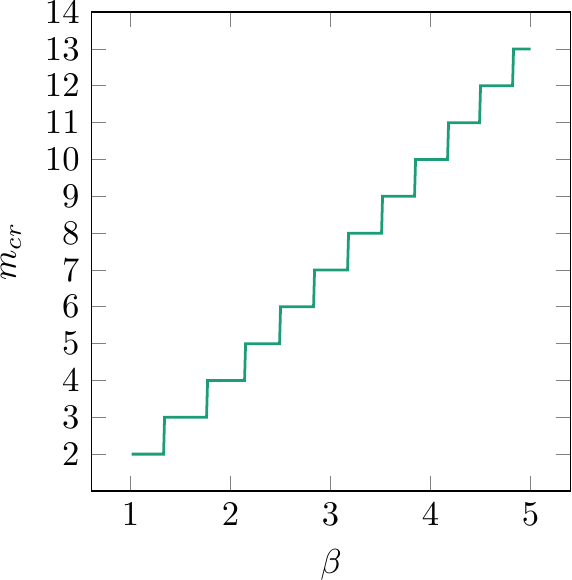}
			\label{fig:Cm1}
			}
		\caption{Marginal stability curves for the residually stressed sphere where $f(R)$ is given by the Eq.~\eqref{eqn:fpolinomiale}, showing the critical $\alpha$ vs. the wavenumber $m$ (left) and  the critical wavenumber $m_{cr}$ vs. $\beta$ (right).}
		\label{fig:C1}
\subfloat{\includegraphics[height=0.35\textwidth]{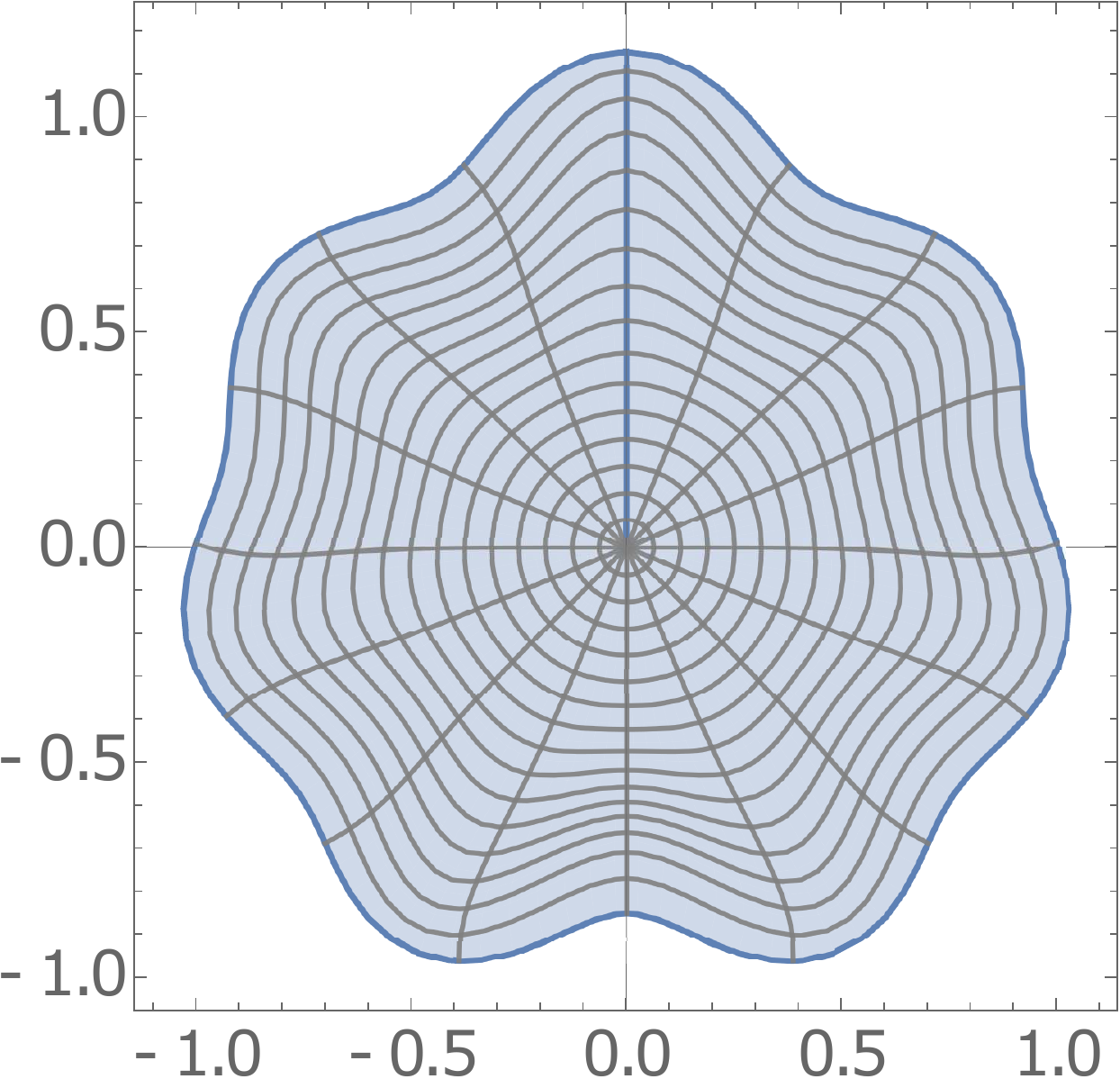}}
\qquad\qquad
\subfloat{\includegraphics[height=0.35\textwidth]{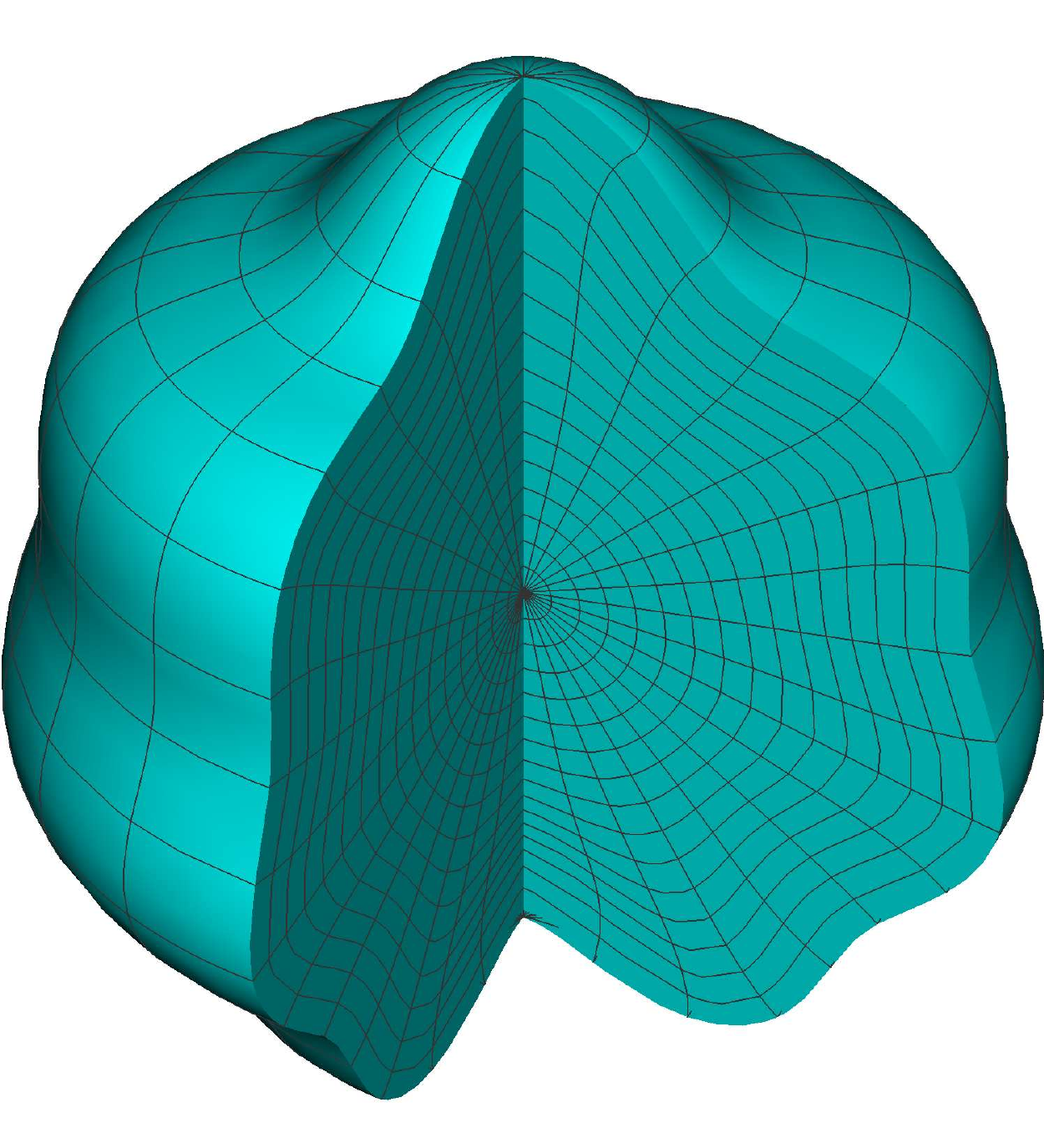}}
\caption{Solution of the linearized incremental problem for $\beta=3$ and $m=m_{cr}=7$ where $f(R)$ is given by the Eq.~\eqref{eqn:fpolinomiale}. The amplitude of the incremental deformation has been arbitrarily set $0.15\,R_o$ for the sake of graphical clarity.}
\label{fig:disp_lin_pol}
\end{figure}

Let us first consider the case in which the expression of $f(R)$ is the exponential polynomial given by Eq.~\eqref{eqn:fpolinomiale}. We use the initial condition given by Eq.~\eqref{eqn:Zinizint}.

We find out that the stop condition given by Eq.~\eqref{eqn:stopconditionint} is satisfied only for negative values of $\alpha$, namely we can find an instability only if the hoop residual stress is tensile close to the center and compressive near the boundary of the sphere.  Moreover, the results are independent on the choice of the $\delta$ in Eq.~\eqref{eqn:centralimp}.

For fixed $\beta$ and $m$, let $\alpha_m$ be the first value such that the stop condition Eq.~\eqref{eqn:stopconditionint} is satisfied, we define the critical wavenumber $m_{cr}$ as the wavenumber with minimum $|\alpha_{m}|$ and we denote such a critical value with $\alpha_{cr}$. In Fig.~\ref{fig:C1} (left)  we depict several marginal stability curves for various $\beta$ whilst in Fig.~\ref{fig:C1} (right)  we plot the critical wavenumber vs. $\beta$. We highlight that, as we increase the parameter $\beta$, the critical wavenumber $m_{cr}$ also increases with a nearly linear behavior.

In Fig.~\ref{fig:disp_lin_pol} we plot the solution of the linearized incremental problem for $\beta=3$ where $m=m_{cr}=7$ (see Fig.~\ref{fig:C1} (right)) and we observe that wrinkles appear in the outer shell of the sphere, where the hoop residual stress is compressive.

\subsubsection{Case (b): logarithmic case}

\begin{figure}[t!]
\centering
\subfloat{\includegraphics[scale=0.85]{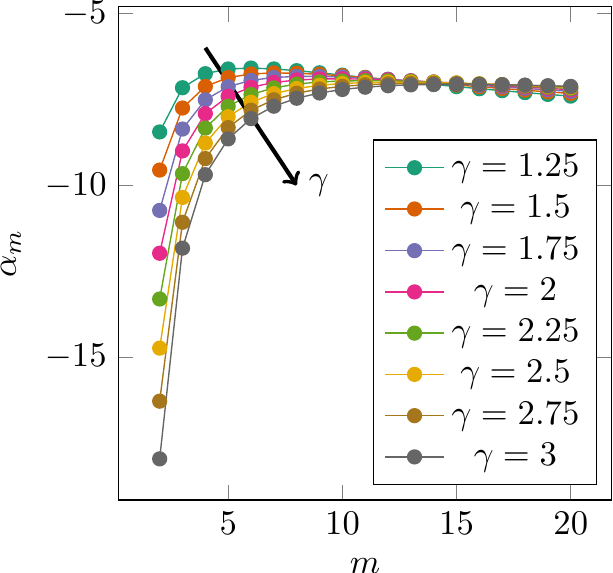}
			\label{fig:Ca2}
			}
		\subfloat{
			\includegraphics[scale=0.85]{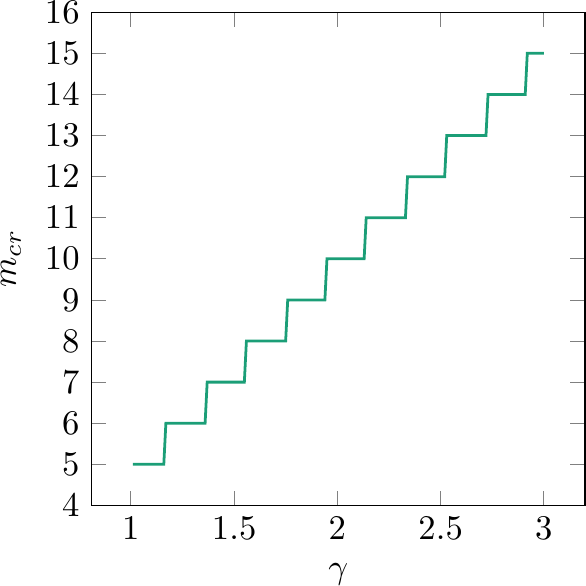}
			\label{fig:Cm2}
			}
		\caption{Marginal stability curves for the residually stressed sphere where $f(R)$ is given by the Eq.~\eqref{eqn:flog}, showing the critical positive $\alpha$ vs. the wavenumber $m$ (left) and the critical wavenumber $m_{cr}$ vs. $\gamma$ (right).}
		\label{fig:C2}
\centering
\subfloat{\includegraphics[height=0.35\textwidth]{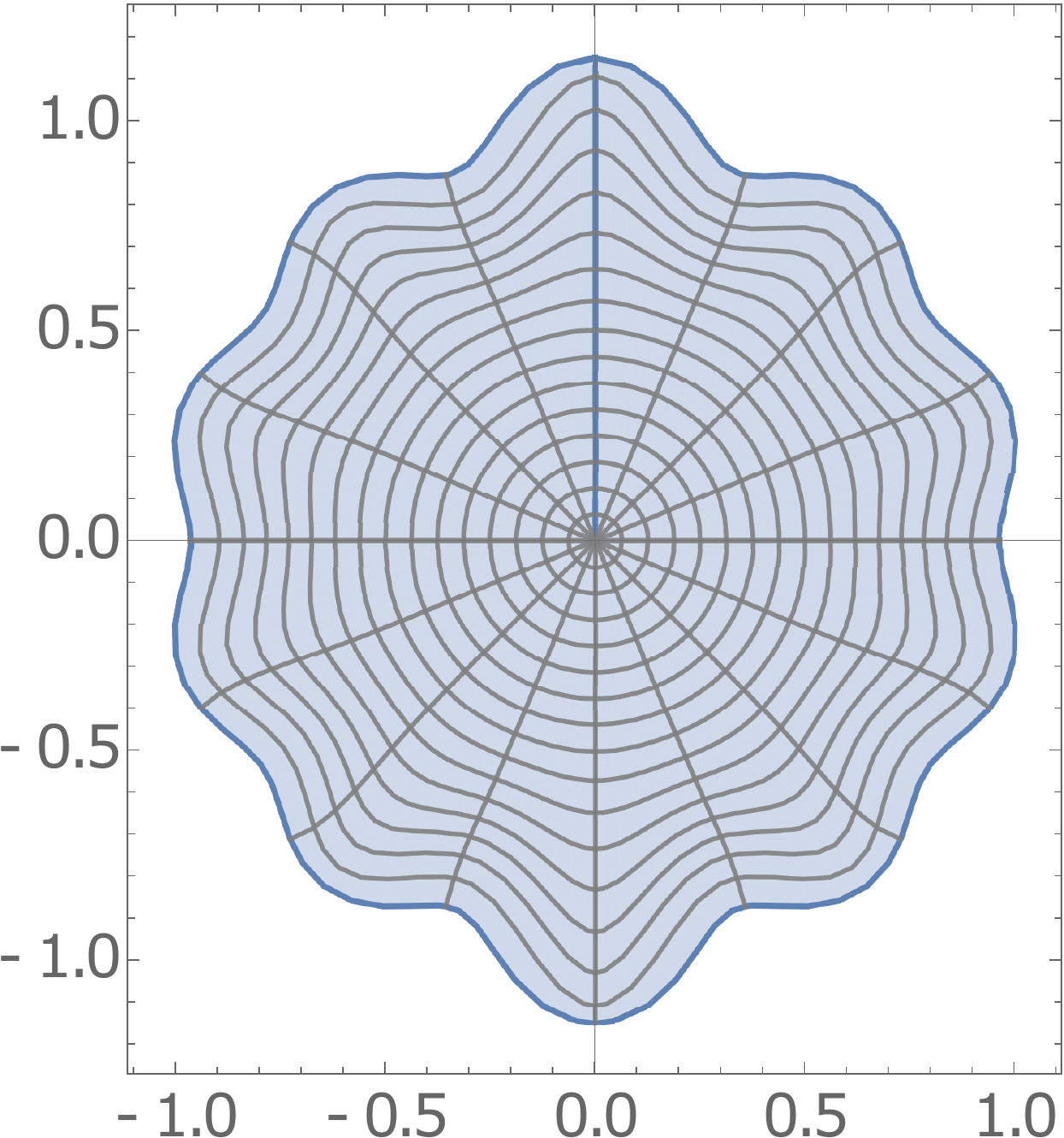}}
\qquad\qquad
\subfloat{\includegraphics[height=0.35\textwidth]{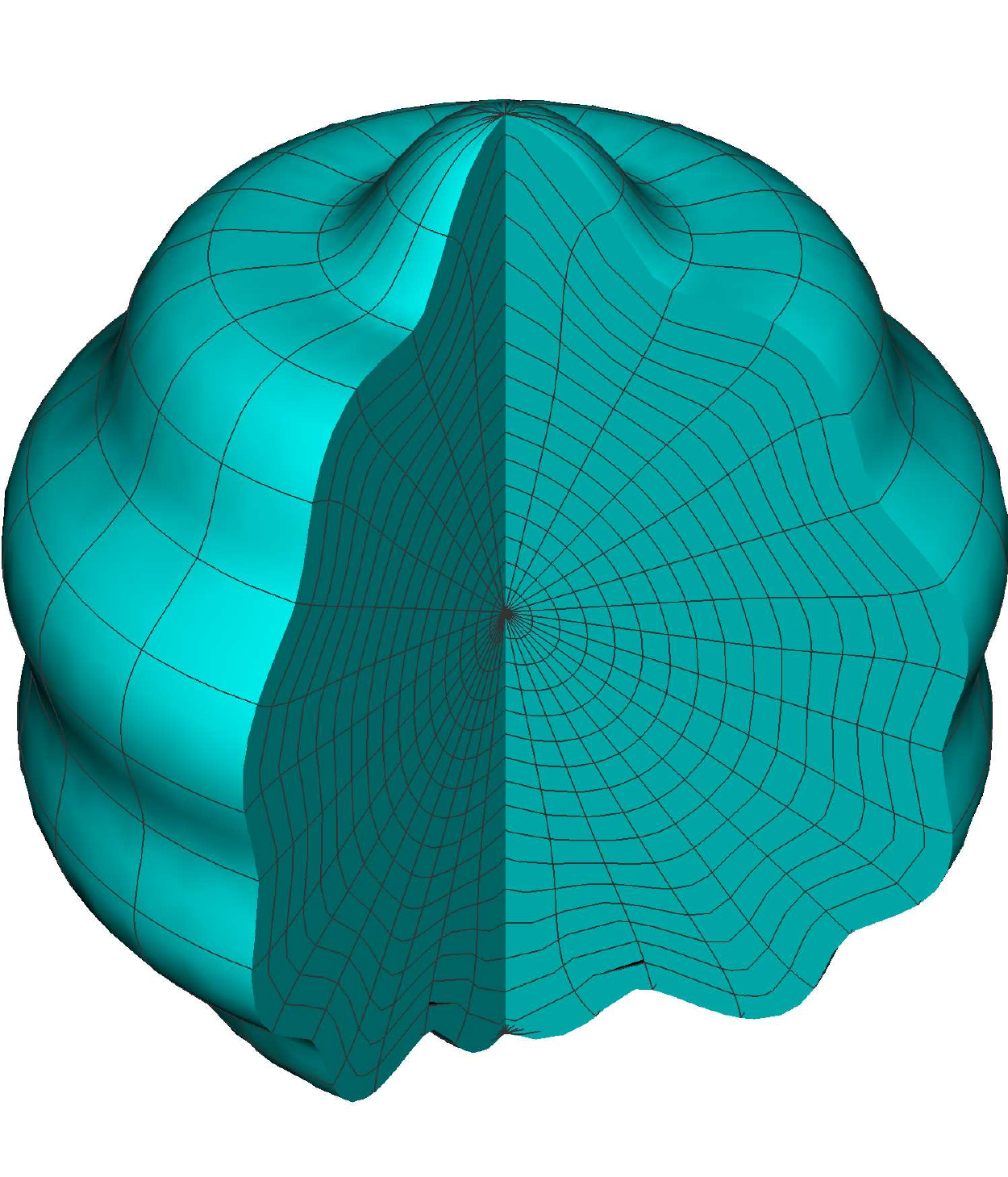}}
\caption{Solution of the linearized incremental problem for $\gamma=2$ and $m=m_{cr}=10$ where $f(R)$ is given by the Eq.~\eqref{eqn:flog}. The amplitude of the incremental deformation has been arbitrarily set $0.15\,R_o$ for the sake of graphical clarity.}
\label{fig:disp_lin_log}
\end{figure}

\begin{figure}[t!]
	\centering
	\subfloat{
		\includegraphics[scale=0.85]{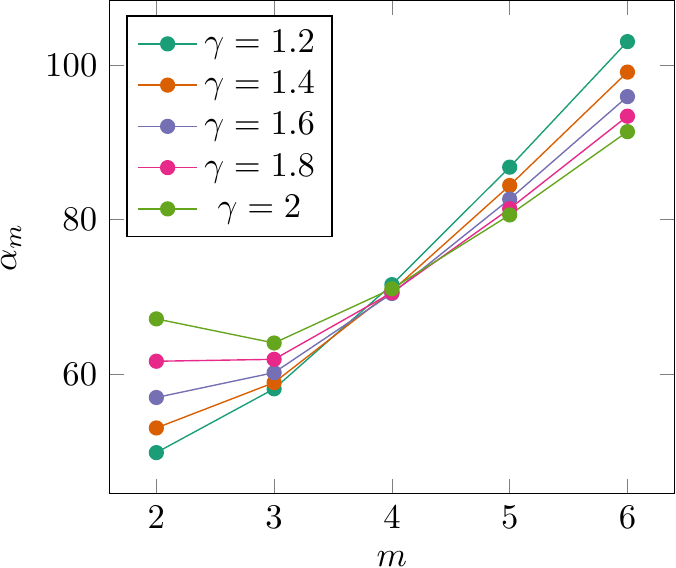}
		\label{fig:Ca3}
	}
	\subfloat{
		\includegraphics[scale=0.85]{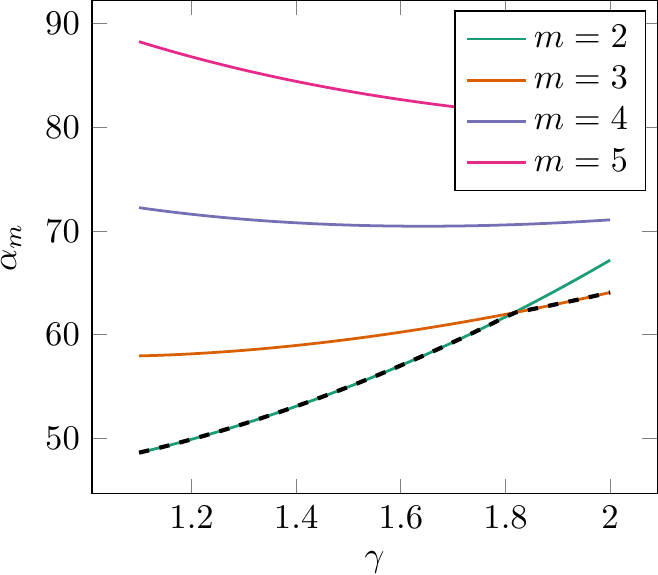}
		\label{fig:Cm3}
	}
	\caption{Marginal stability curves for the residually stressed sphere where $f(R)$ is given by the Eq.~\eqref{eqn:flog}, showing $\alpha_m$ vs. the wavenumber $m$ (left) and $\gamma$ (right). The black dashed curves on the right is the plot of the $\alpha_{cr}$ vs. $\gamma$.}
	\label{fig:C3}
	\centering
	\subfloat{\includegraphics[height=0.35\textwidth]{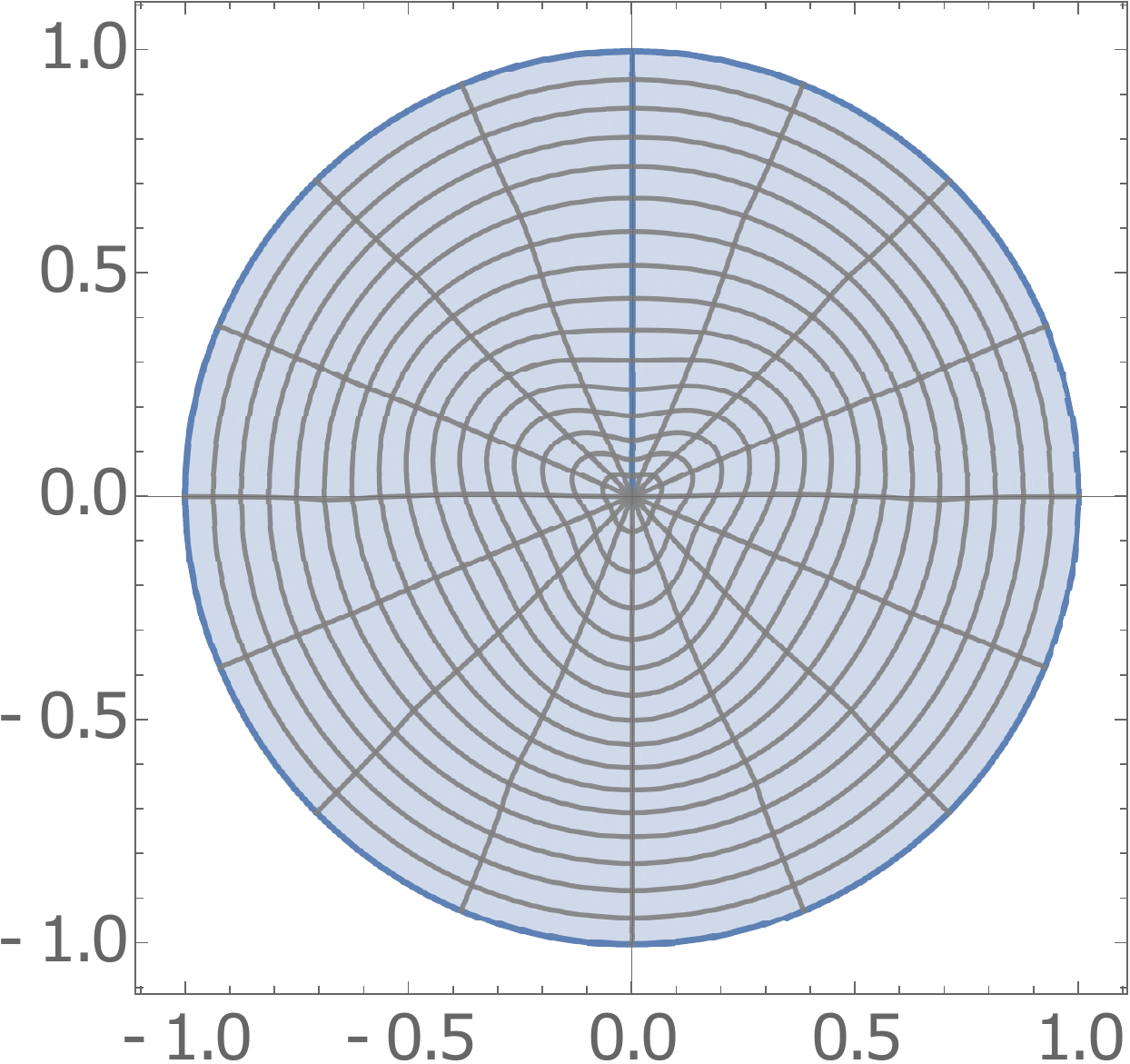}}
\qquad\qquad
	\subfloat{\includegraphics[height=0.35\textwidth]{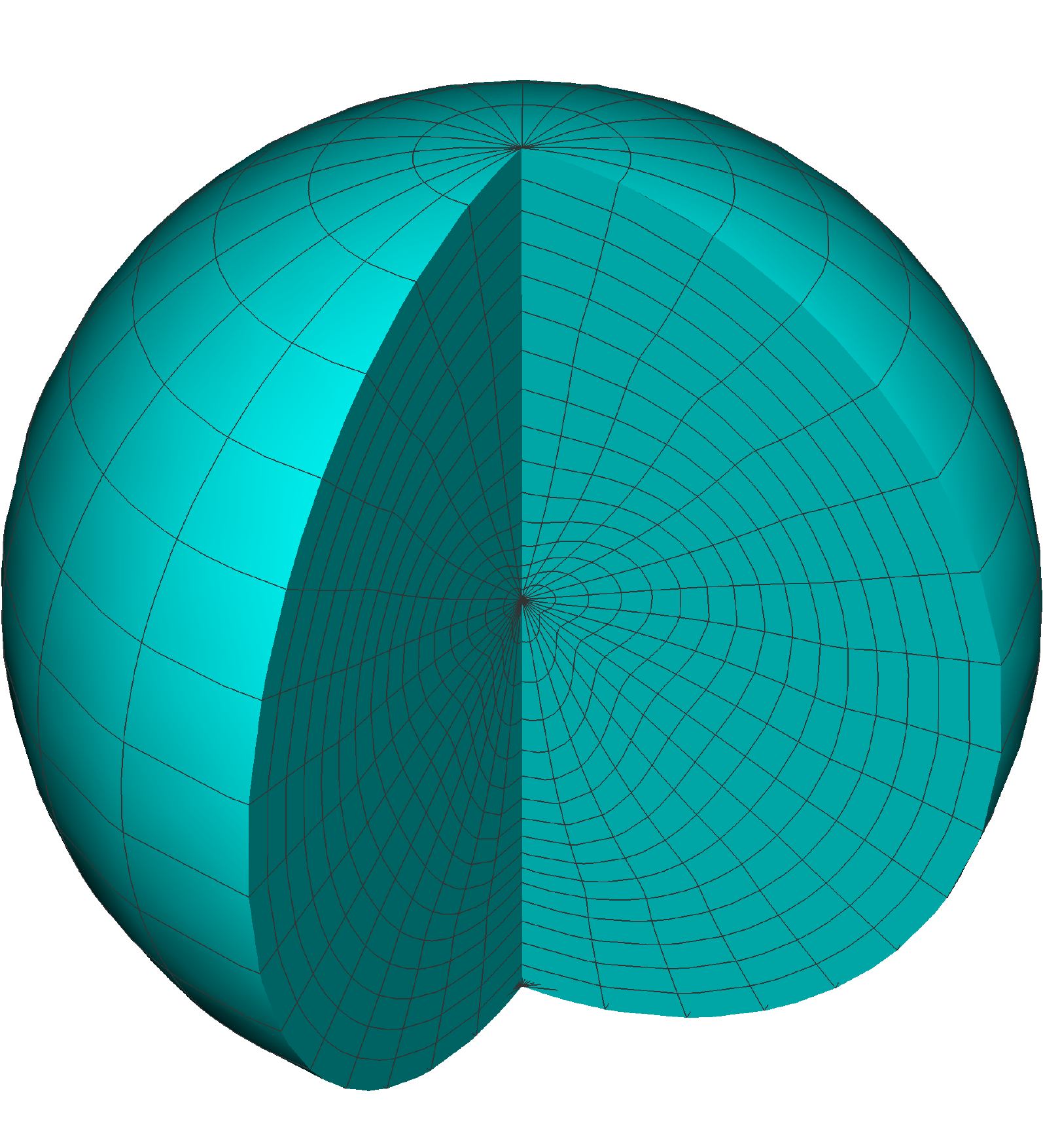}}
	\caption{Solution of the linearized incremental problem for $\gamma=2$ and $m=m_{cr}=3$ where $f(R)$ is given by the Eq.~\eqref{eqn:flog}. The amplitude of the incremental deformation has been arbitrarily set $0.15\,R_o$ for the sake of graphical clarity.}
	\label{fig:disp_lin_log_pos}
\end{figure}

Let us now consider the case in which $f(R)$ is given by Eq.~\eqref{eqn:flog}. We find that the residually stressed sphere is unstable for both positive and negative values of $\alpha$.

When we consider positive values for the control parameter $\alpha$, we integrate the differential Riccati equation given by Eq.~\eqref{eqn:Riccati} from $R=R_o$, using the initial condition given by Eq.~\eqref{eqn:Zinizext}, and using the stop condition at $R=R_c$ given by Eq.~\eqref{eqn:stopconditionext}.

On the other hand, when $\alpha$ is negative, we use as the initial condition the Eq.~\eqref{eqn:Zinizint} and as stop condtion the Eq.~\eqref{eqn:stopconditionint}. This means that we integrate the Riccati equation from the interior to the exterior.

Let us first consider the case in which $\alpha$ is negative, namely when the hoop stress is compressive at the boundary (see Fig.~\ref{fig:res_stress}). In this framework in Fig.~\ref{fig:C2} (left),  we depict several marginal stability curves for various $\gamma$, whereas in Fig.~\ref{fig:C2} (right)  we plot the values of the critical wavenumber vs. the parameter $\gamma$. As previously observed, by increasing $\gamma$, also the critical wavenumber $m_{cr}$ increases with a nearly linear dependence.

In Fig.~\ref{fig:disp_lin_log} we plot the solution of the linearized incremental for $\gamma=2$, where $m=m_{cr}=7$ (see Fig.~\ref{fig:C2}, right); as in the polynomial case, we can notice how wrinkles appear in the outer rim of the domain, where the hoop residual stress is compressive.

We perform the same calculations for the case in which $\alpha$ is positive. In Fig.~\ref{fig:C3} we depict the resulting marginal stability curves for various $\gamma$ and $m$.

In Fig.~\ref{fig:disp_lin_log_pos} we plot the solution of the linearized incremental problem for $\gamma=2$ and $m=m_{cr}=3$. We highlight that the displacement is localized in the center of the sphere whereas the exterior part remains almost undeformed.

Also in this case, we found that all the results exposed are independent of the chosen value of $\delta$ in Eq.~\eqref{eqn:centralimp}.

In the next section, we implement a finite element code in order to investigate the fully non-linear evolution of the morphological instability.

\section{Finite element implementation and post-buckling analysis}

\subsection{Mixed finite element implementation}

\begin{figure}[b!]
\centering
\includegraphics[width=0.7\textwidth]{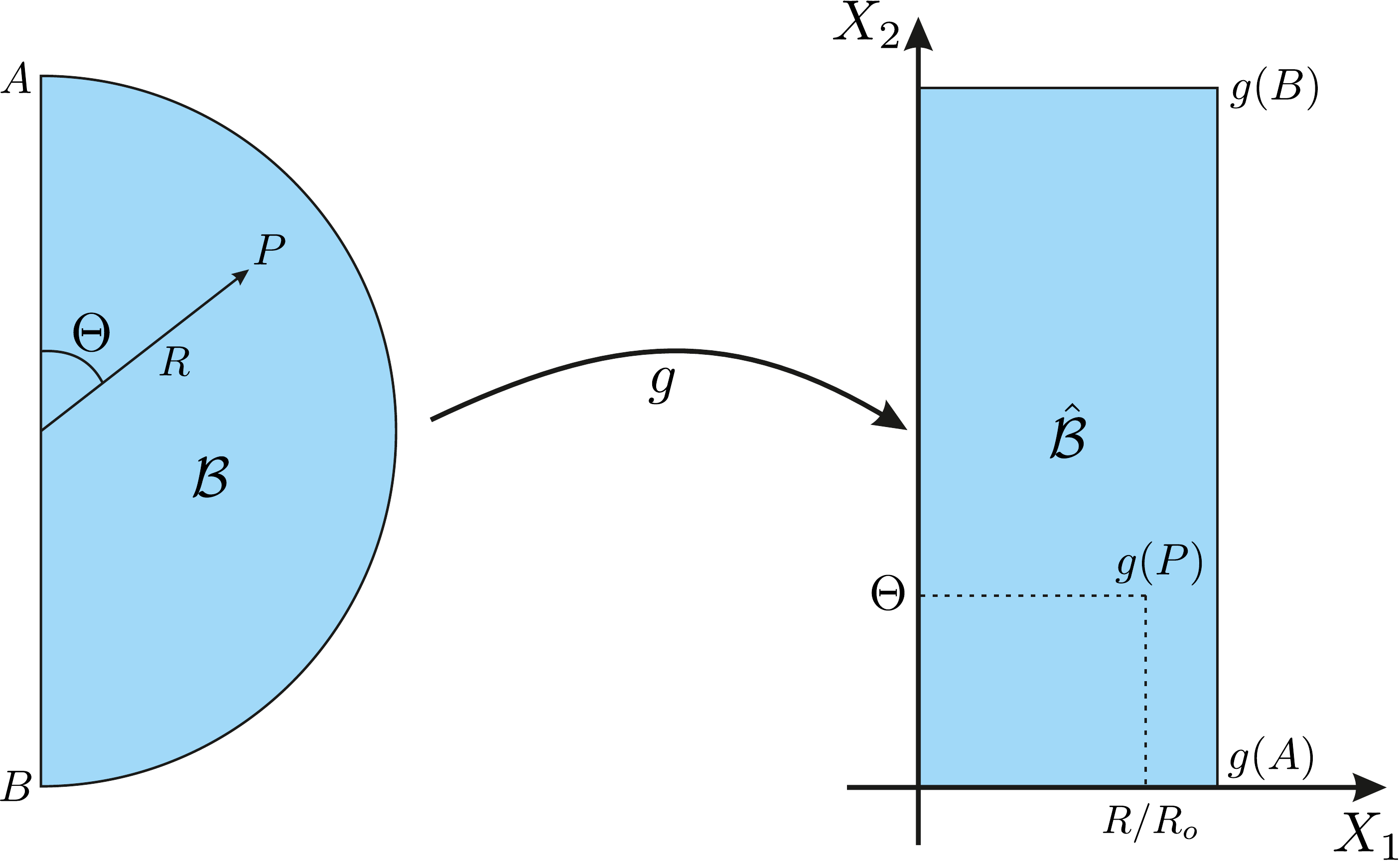}
\caption{Representation of the conformal mapping between the physical domain  $\mathcal{B}$ and its conformal image $\hat{\mathcal{B}}$, defined through the coordinate transformation in Eq. \eqref{CM}.}
\label{fig:trasf}
\end{figure}

We use a mixed variational formulation of the problem implemented with the open source project FEniCS~\cite{logg2012automated}. Let $\mathcal{B}$ be a semicircle and $\hat{\mathcal{B}}=(0,1)\times(0,\,\pi)$ as depicted in Fig. \ref{fig:trasf}. We define $g:\mathcal{B}\rightarrow\hat{\mathcal{B}}$ as the mapping that associates each point in $\mathcal{B}$ with the point in $\R^2$ such that the two components are the normalized radial distance $R/R_o$ and the polar angle $\Theta$. Hence, denoting by $X_1$ and $X_2$ the first and the second coordinates respectively and by $\vect{e}_1$ and $\vect{e}_2$ the canonical unit basis vectors, we get that
\begin{equation}
\label{CM}
X_1=\frac{R}{R_o},\qquad X_2=\Theta.
\end{equation}

We solve the nonlinear problem using a triangular mesh $\hat{\mathcal{B}}_h$ obtained through the discretization of the set $\hat{\mathcal{B}}$. The mesh is composed of $14677$ elements, $7519$ vertices and the maximum diameter of the cells is $0.033$.

We use the Taylor--Hood elements $\vect{P}_2$-$P_1$, discretizing the displacement field by using piecewise quadratic functions and the pressure field by piecewise linear functions.
The Taylor-Hood element is numerically stable for linear elasticity problems  \cite{boffi2013mixed} and has been used in several applications  of non-linear elasticity \cite{auricchio2005stability}.

In order to study the behavior of the bifurcated solution in the post-buckling regime, we impose a small imperfection on the mesh at the boundary \cite{ciarletta2016morphology} with the form given by Eqs.~\eqref{eqn:uPm}-\eqref{eqn:vPm}, where $m$ is the critical wavenumber obtained from the linear stability analysis and the amplitude is of the order of $10^{-4}$.

We impose as boundary conditions
\begin{equation}
\label{eqn:BCdiscr}
\left\{
\begin{aligned}
&\vect{u}_h=\vect{0} &&\text{if }X_1=0,\\
&\vect{u}_h\cdot\vect{e}_2=0\text{ and }\vect{e}_1\cdot\tens{S}^T_h\vect{e}_2=0 &&\text{if }X_2=0\text{ or }X_2=\pi,\\
&\tens{S}^T_h\vect{e}_1=\vect{0} &&\text{if }X_1=1;\\
\end{aligned}
\right.
\end{equation}
where $\vect{u}_h$ is the discretized displacement field and $\tens{S}_h$ the discretized first Piola--Kirchhoff stress tensor.

The problem is solved by using an iterative Newton--Raphson method whilst adaptively  incrementing the control parameter $\alpha$. The code automatically adjusts the increment of this parameter either near the marginal stability threshold or when the Newton method does not converge.

Each step of the Newton--Raphson method is performed using PETSc as a linear algebra back-end and then the linear system is solved through an LU decomposition.


\subsection{Results of the finite element simulations}

\subsubsection{Case (a): exponential polynomial case}

\begin{figure}[b!]

\subfloat{
		\includegraphics[scale=0.85]{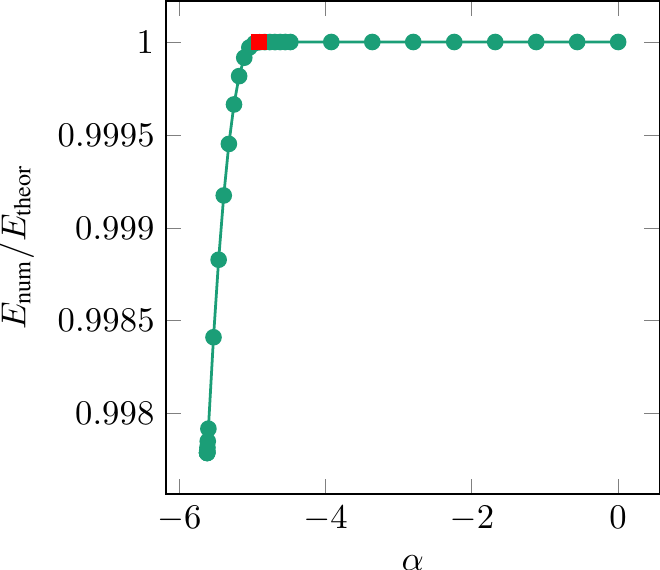}
		\label{fig:enratio}
	}
	\subfloat{
		\includegraphics[scale=0.85]{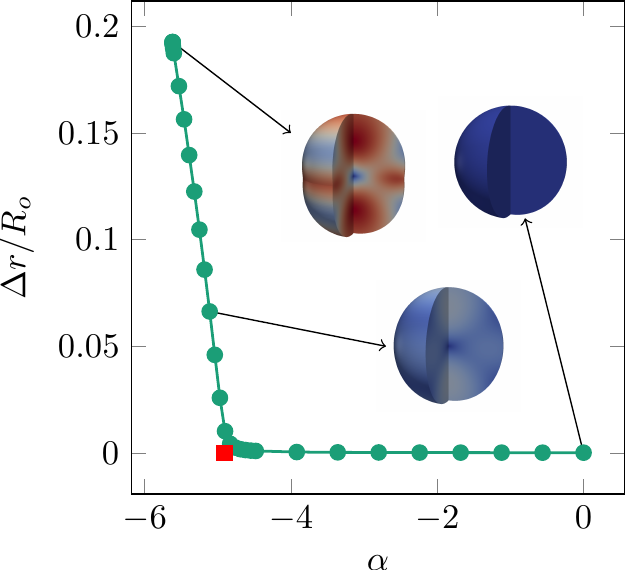}
		\label{fig:amplitude}
	}
	\caption{Plots of the ratio $E_\text{num}/E_\text{theor}$ (left) and  the normalized buckling amplitude $\Delta r/R_o$ (right) versus the control parameter $\alpha$. The numerical results are in good agreement with the theoretical instability threshold $\alpha_{cr}=-4.9084$ (red square marker).
	\label{fig:amplitudetheo}}
\end{figure}
\begin{figure}[t!]
\centering
\includegraphics[scale=0.85]{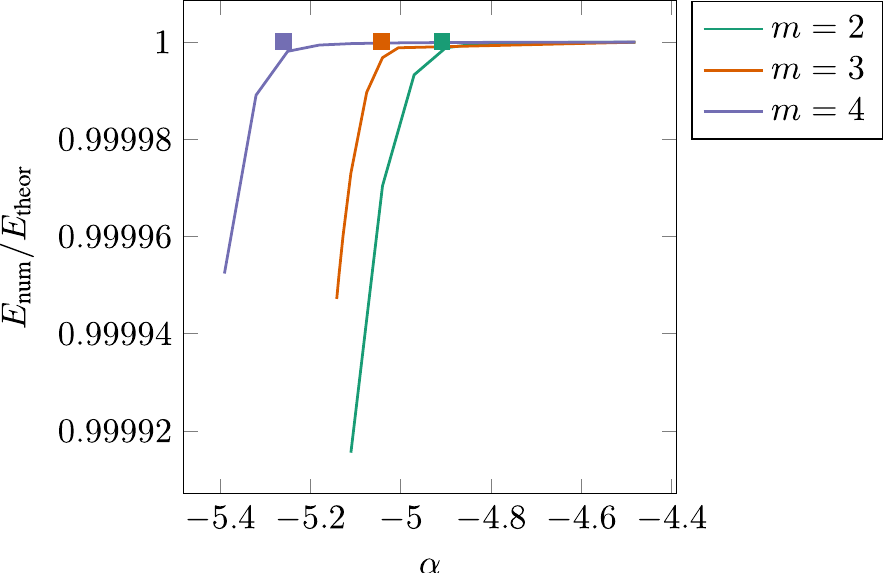}
		\caption{Comparison between the ratios $E_\text{num}/E_\text{theor}$ vs. the wavenumber $m$. The squares denote the thresholds $\alpha_m$ computed in the previous section.}
		\label{fig:enratio_comparison}
\end{figure}

\begin{figure}[tb!]
\centering
\subfloat{\includegraphics[width=0.4\textwidth]{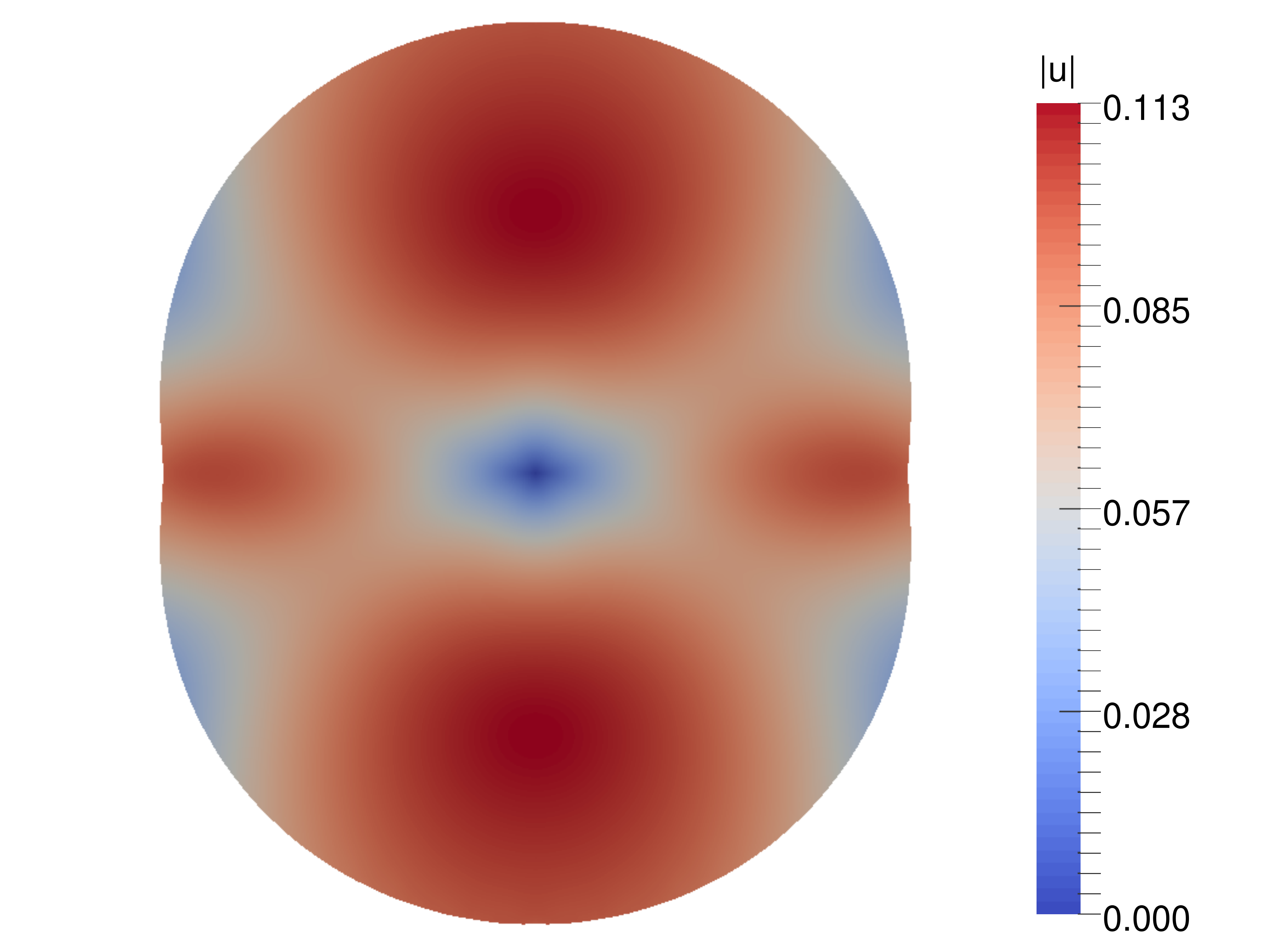}}
\subfloat{\includegraphics[width=0.4\textwidth]{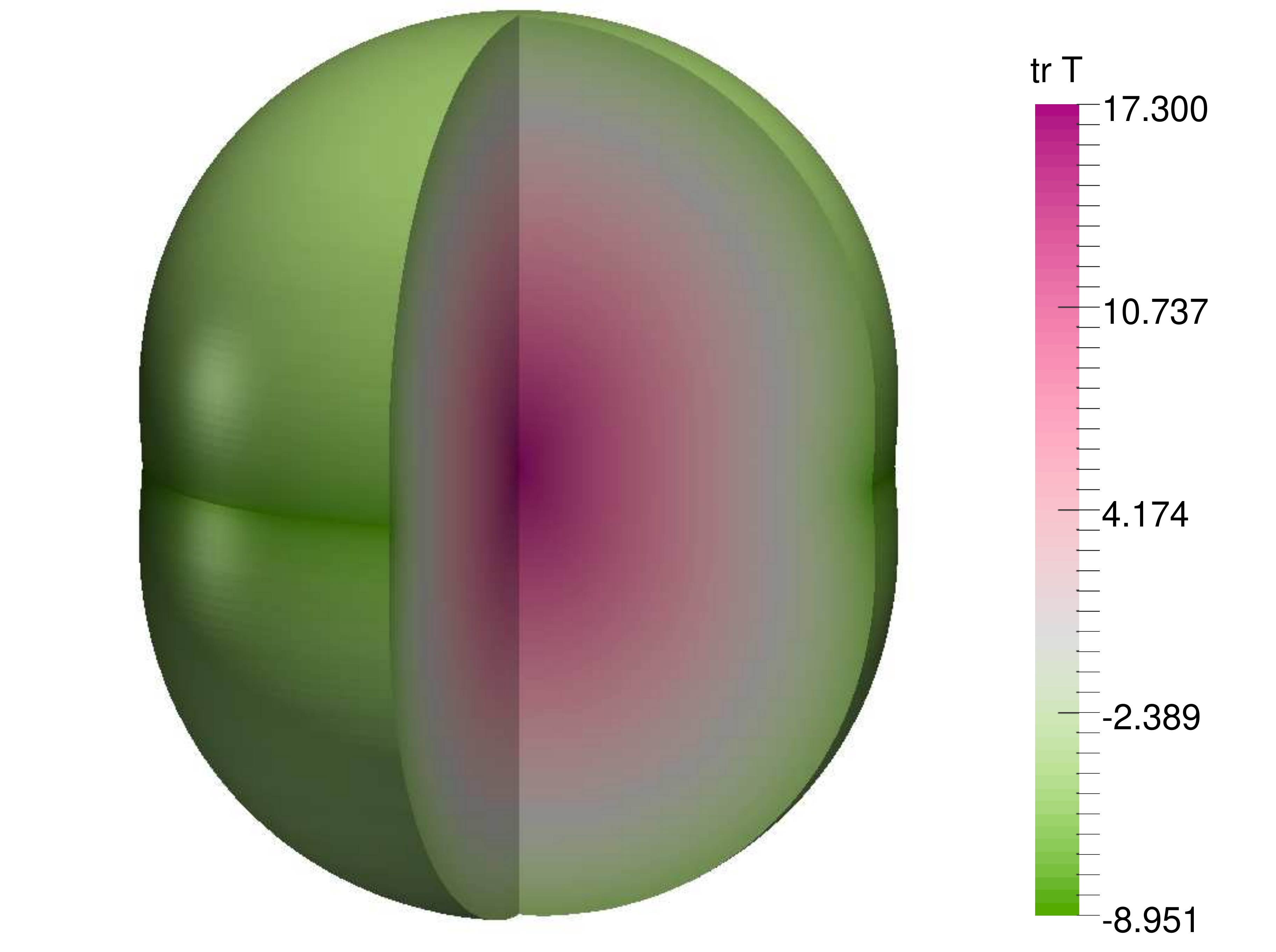}}\\
\subfloat{\includegraphics[width=0.4\textwidth]{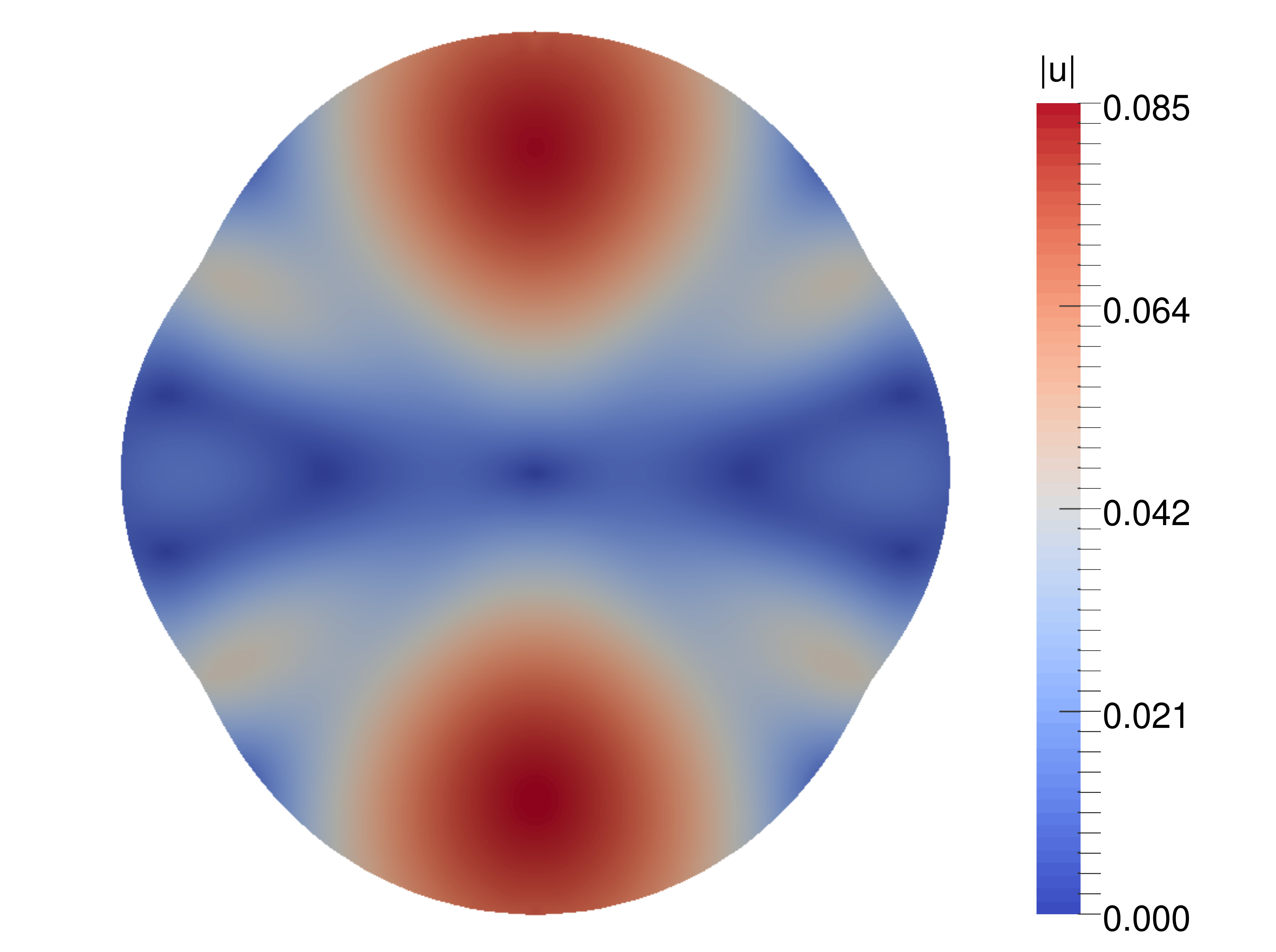}}
\subfloat{\includegraphics[width=0.4\textwidth]{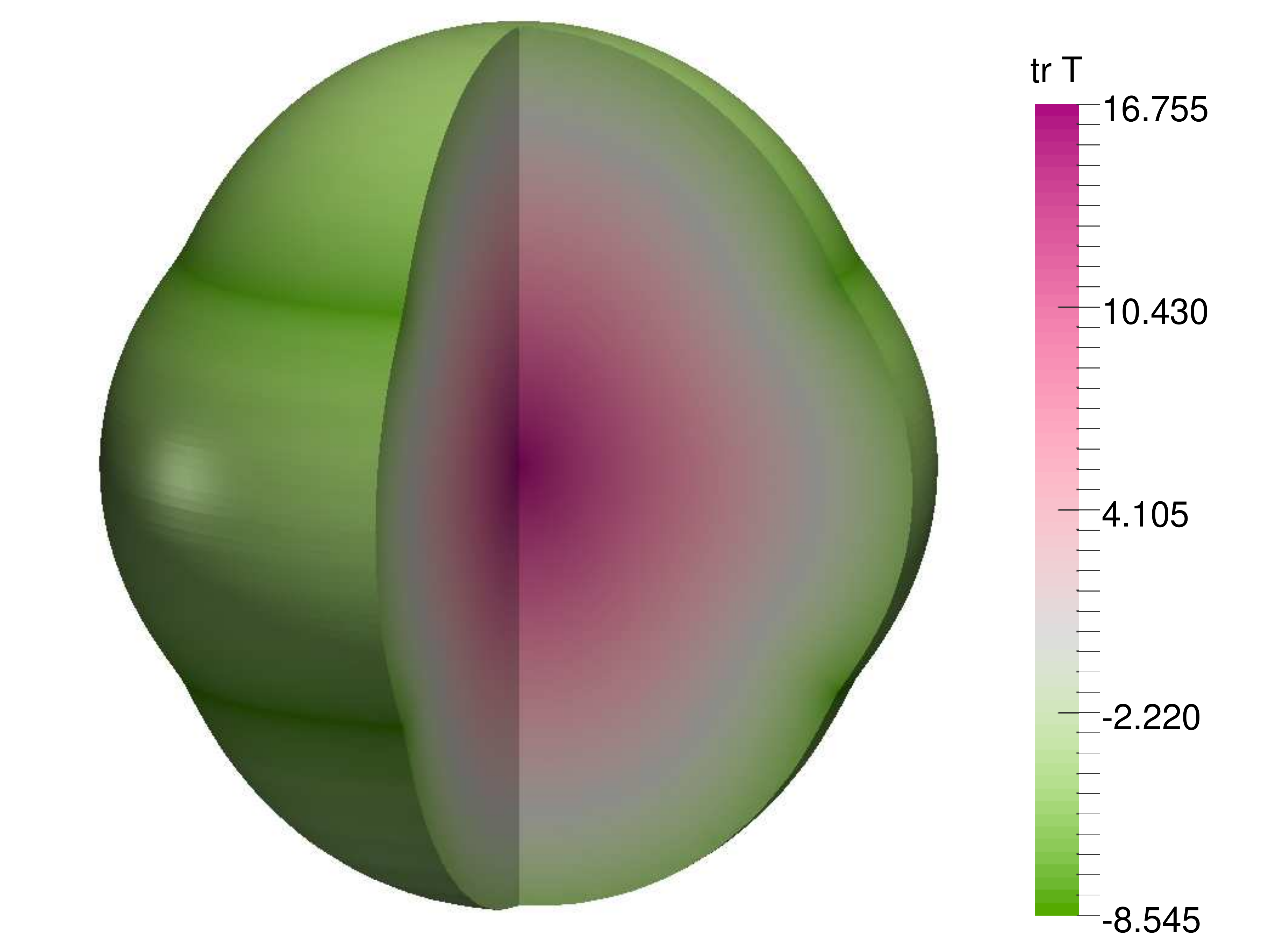}}
\caption{Plot of the deformed configuration when $f(R)$ is given by Eq.~\eqref{eqn:fpolinomiale}, $\beta=1.1$, $\alpha=-5.62$ and $m=m_\text{cr}=2$ (top); $\alpha=-5.55$ and $m=4$ (bottom). The color bars indicate the norm of the displacement $\|\vect{u}_h\|$ (left) and the trace of the Cauchy stress tensor normalized with respect to the shear modulus $\mu$ (right). On the right we depict a $3$D representation of the deformed sphere.}
\label{fig:beta11}
\end{figure}

We first show the results for the case in which $f(R)$ is given by Eq.~\eqref{eqn:fpolinomiale}. We denote by $E_\text{num}$ the total strain energy of the deformed material, and by $E_\text{theor}$ the theoretically computed strain energy of the undeformed sphere, namely in the reference configuration. We remark that the strain energy density in the undeformed reference configuration may not be zero. Indeed, setting $\tens{F}=\tens{I}$ in \eqref{eqn:strainenergyexpr}, it is easy to check  that the energy density vanishes only if $\tens{\Sigma}=\tens{0}$.  Thus, the presence of pre--stresses  is physically related to the fact that some mechanical energy is already stored inside the material.

In Fig.~\ref{fig:amplitudetheo} (left) we plot the ratio between $E_\text{num}$ and $E_\text{theor}$ vs. $\alpha$ when $\beta=1.1$; the mode of the imperfection applied on the mesh is the critical one $m_\text{cr}=2$, we also computed the amplitude of the pattern, defined as 
\[
\Delta r\coloneqq\max_{\Theta\in[0,\pi]}r_h(R_o,\,\Theta)-\min_{\Theta\in[0,\pi]}r_h(R_o,\,\Theta),
\]
where $r_h$ is the discretized deformation field in the radial direction (Fig.~\ref{fig:amplitudetheo} (right)). We observe that there is a smooth increase of such an amplitude when the control parameter is lower than $\alpha_{cr}$.
When performing a cyclic variation of the control parameter, decreasing $\alpha$ first and then increasing it to zero, both the amplitude of the wrinkling and the energy ratio do not encounter any discontinuity and they both follow the same curve in both directions.

Since $\alpha_{cr}$ is very close to the other values $\alpha_m$, in Fig.~\ref{fig:enratio_comparison} we compare the energy ratio also for the cases in which the wavenumber of the imperfection is not the critical one, specifically  $m=3$ and $m=4$. We can observe that there is a continuous decrease of such a ratio when the threshold $\alpha_{m}$ is reached. From the picture we can also notice that there is no intersection of the curves that represent the ratio of the energies, thus suggesting the absence of secondary bifurcations.

Setting  $\beta=1.1$, in Fig.~\ref{fig:beta11} we depict the deformed configuration of the sphere when $\alpha=-5.62$, when $m=m_\text{cr}=2$ (top) and $\alpha=-5.55$ when $m=4$ (bottom), with the color bar we indicate the norm of the displacement $\|\vect{u}_h\|$ (left) and the trace of the Cauchy stress tensor $\tens{T}_h$ normalized with respect to the shear modulus $\mu$ (right).

\begin{figure}[b!]
\centering
\subfloat{\includegraphics[scale=0.85]{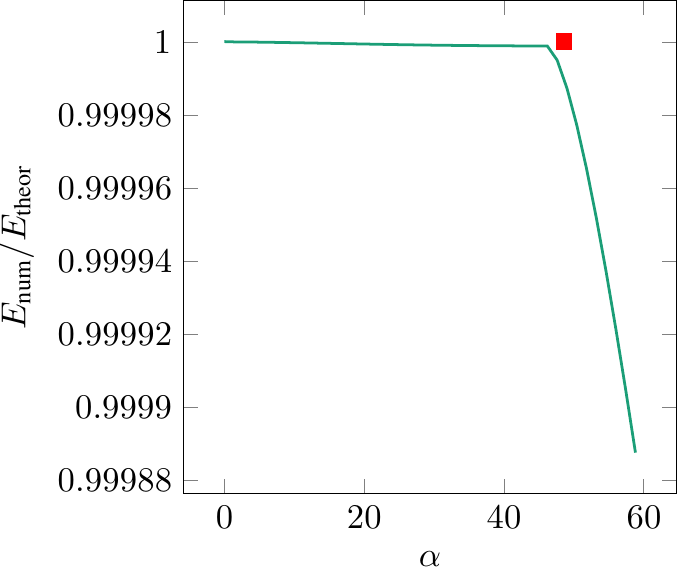}}
		\subfloat{\includegraphics[scale=0.85]{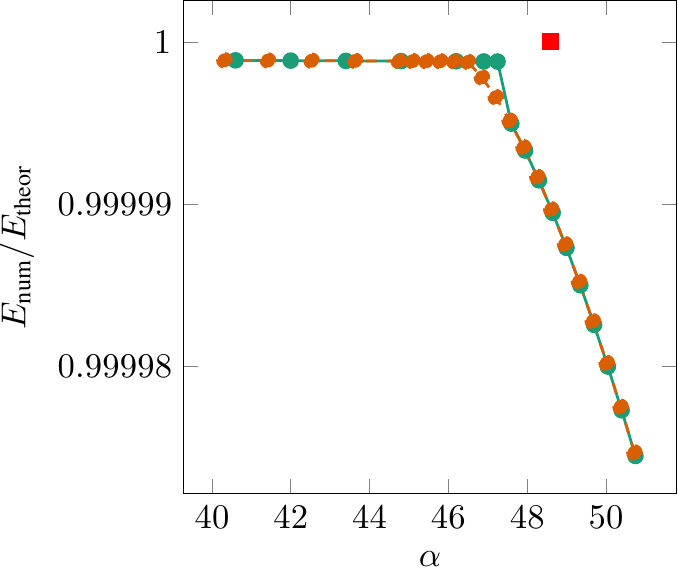}}
		\caption{Ratio between $E_\text{num}/E_\text{theor}$ versus the control parameter $\alpha$(Left) when $f(R)$ is given by Eq.~\eqref{eqn:flog} for $\gamma=1.1$ and $m=m_{cr}=2$. We performed a cyclic variation of the control parameter $\alpha$ (right), first increasing it beyond the linear stability threshold  (green solid line) and then decreasing it down to the initial value (orange dashed line). In both plots, the red squares denote the threshold $\alpha_2=48.60$ computed in the previous section.}
		\label{fig:enratio_comparison_log}
\end{figure} 

\begin{figure}[t!]
\centering
\subfloat{\includegraphics[width=0.45\textwidth]{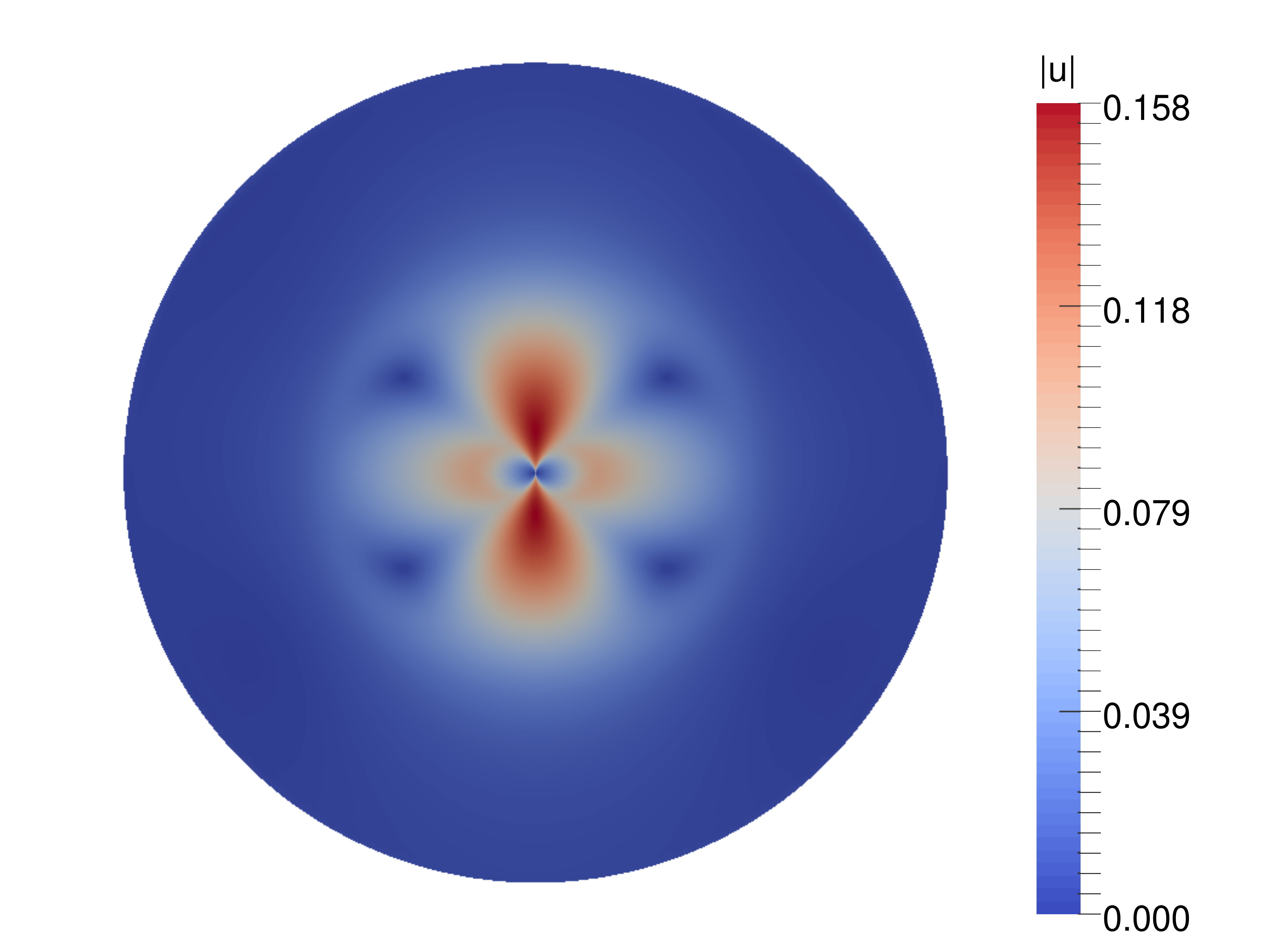}}
\subfloat{\includegraphics[width=0.45\textwidth]{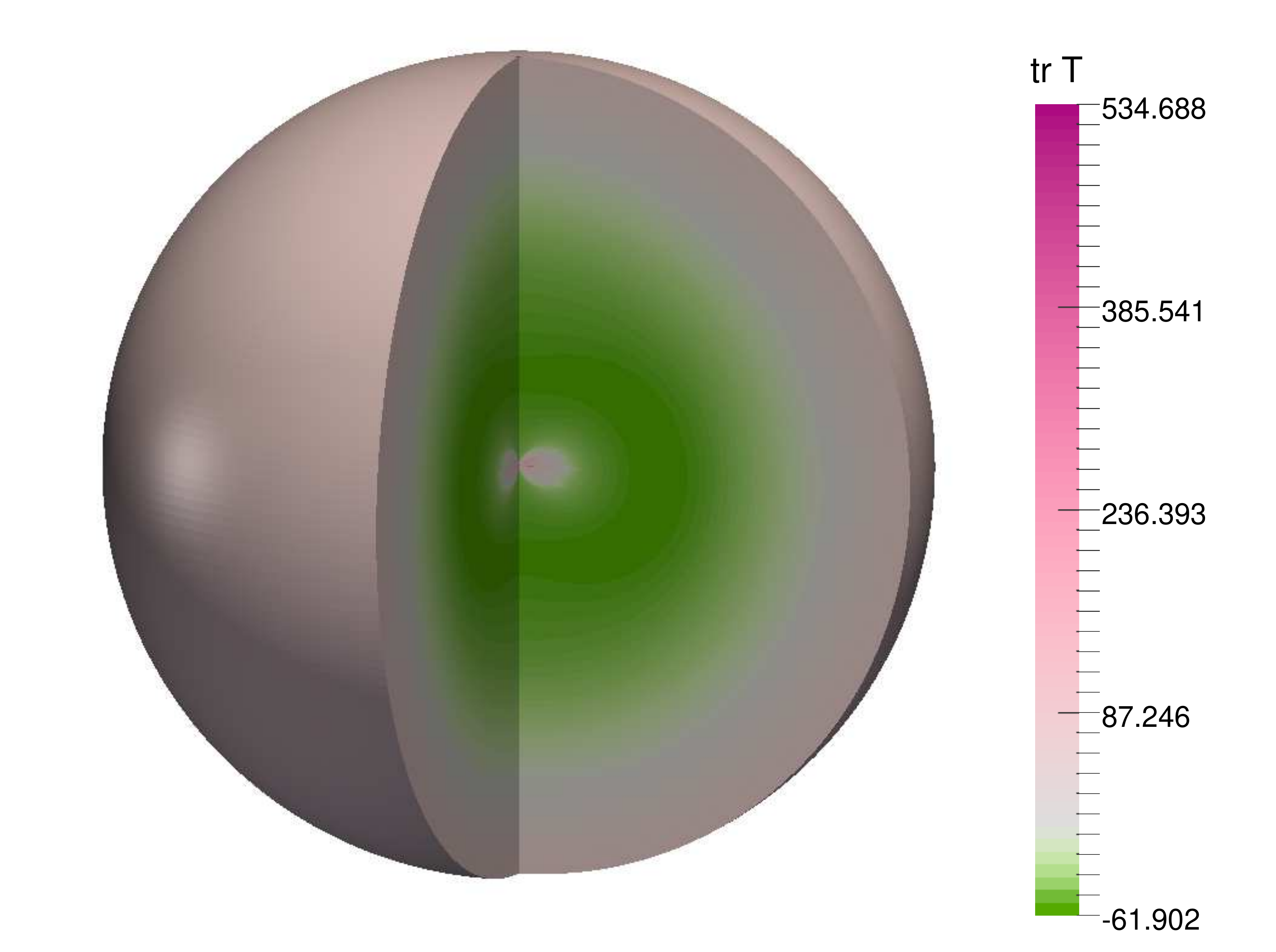}}\\
\caption{Plot of the deformed configuration when $f(R)$ is given by Eq.~\eqref{eqn:flog}, $\gamma=1.1$, $\alpha=58.8$ and $m=m_\text{cr}=2$. The color bars indicate the norm of the displacement $\|\vect{u}_h\|$ (left) and the trace of the Cauchy stress tensor (right). On the right we depict a $3$D representation of the deformed sphere.}
\label{fig:betalog11}
\end{figure}

\subsubsection{Case (b): logarithmic case}

We performed the same numerical procedure for simulating  the logarithmic case.

We considered the case in which $\alpha$ is positive. From the linear stability analysis we expect that the instability is localized in the interior part of the sphere (Fig.~\ref{fig:disp_lin_log_pos}).

Let $\gamma=1.1$, in Fig.~\ref{fig:enratio_comparison_log} we plot the ratio $E_\text{num}/E_\text{theor}$ at varying $\alpha$. We performed a cyclic variation of the control parameter $\alpha$, first increasing it and then decreasing it down to zero Fig.~\ref{fig:enratio_comparison_log} (right).  We highlight the presence of both a jump across the linear threshold and hysteresis, thus highlighting the presence of a subcritical  bifurcation.
The linear stability threshold is in good agreement with the theoretical prediction, given that subcritical bifurcations have a higher sensitivity to imperfection than supercritical ones.

In Fig.~\ref{fig:betalog11} we show the deformed configuration of the sphere when $\alpha=58.8$ for $\gamma=1.1$, where the color bars  indicate the norm of the displacement $\|\vect{u}_h\|$ and the trace the Cauchy stress tensor $\tens{T}_h$ normalized with respect to the shear modulus $\mu$. 

We remark that we obtain small numerical oscillations of the displacement field near the center of the sphere in the fully nonlinear post-buckling regime. These errors eventually  get amplified during the computation of the stress field, and the numerical solution no longer converges. In some cases, we observed that the Newton method failed to converge for some different values of the parameter $\gamma$ when $\alpha$ is just beyond the marginal stability threshold $\alpha_{cr}$. The improvement of the numerical continuation method is outside the scope of this work, but we acknowledge that a different approach, e.g. using scalable iterative solvers and preconditioners  \cite{farrell2017linear}, could improve the stability of the numerical solution in the post-buckling regime. 

\section{Discussion and concluding remarks}

This work investigated the morphological stability of a soft elastic sphere subjected to residual stresses.

In the first part, we modeled the sphere as a hyperelastic material by introducing a strain energy depending explicitly on the deformation gradient and on the initial stress \cite{gower2015initial,gower2016new}. In this way, we can avoid the classical deformation gradient decomposition \cite{rodriguez1994stress} which has the drawback of requiring the a priori knowledge of a virtual relaxed state.

Secondly, we described the residual stress fields by using a function $f(R)$ that denotes the radial component of the residual stress. This function depends on the dimensionless parameters $\alpha,\,\beta$ and $\gamma$, where $\alpha$ is the normalized intensity of the residual stress whereas $\beta$ and $\gamma$ describe the spatial distribution of the residual stress components within the sphere.

We investigate two possible distributions of the radial residual stress $f(R)$, one based on a polynomial function, the other on a logarithmic one. We denote these two choices as case (a) and (b) respectively. 

We performed the linear stability analysis in both cases by using the theory of incremental deformations superposed on the undeformed, pre-stressed configuration.
In order to solve the incremental boundary value problem, we used the Stroh formulation and the surface impedance matrix method to transform it into the differential Riccati equation given by Eq.~\eqref{eqn:Riccati} \cite{norris2012elastodynamics} .

We integrated numerically the resulting incremental initial value problem by  iterating the control parameter until a stop condition is reached, in order to find the marginal stability thresholds. We found out that the morphological transition occurs in the region where the hoop residual stress reaches its maximum magnitude in compression.

In the case (a) we find an instability only for $\alpha<0$, whilst in case (b) we find an instability for both $\alpha$ positive and negative. In this latter case, when $\alpha$ is positive the instability occurs in the inner region of the sphere whereas if $\alpha$ is negative it is localized in the external region. The results of such analysis are reported in Figures ~\ref{fig:C1}-\ref{fig:disp_lin_log_pos}.

Finally, we implemented a numerical procedure by using the mixed finite element method in order to approximate the fully non-linear problem.  After the validation of the numerical simulations obtained by the comparison with the results of the linear stability analysis, we analyzed the resulting morphology in the fully non-linear regime.

In the case (a), the instability is localized in the external part of the sphere where the hoop residual stress is compressive. The continuous transition from the initial configuration to the buckled state indicates that the bifurcation is supercritical.

In the case (b), the instability is localized near the center of the sphere when the parameter $\alpha>0$. In contrast to the previous case, the bifurcation is found to be subcritical, thus suffering a jump across the linear stability threshold. The results of these simulations are reported in Figures ~\ref{fig:amplitudetheo}-\ref{fig:betalog11}.

Future efforts will be directed to improve the proposed analysis either by implementing of a fully 3D numerical model in order to study the secondary bifurcation that might appear in the azimuthal direction or by accounting for the presence of material anisotropy, a major determinant for the residual stress distribution in living matter, e.g.  tumor spheroids \cite{dolega2017cell}. 

In summary,  this work proposes a novel approach that may prove useful guidelines for engineering applications.  For example, it may be of interest for achieving a  nondestructive determination of the pre--stresses in soft spheres. 
 Whilst the currently used method  consists in cutting the material and inferring  the residual stresses through the resulting deformation \cite{stylianopoulos2012causes}, the proposed model explicitly  correlates both the mechanical response of the material and its morphology  with the underlying distribution of pre-stresses. Moreover, the proposed static analysis based on the Stroh formulation can be easily adapted to solve the corresponding elasto--dynamic problem in a solid sphere \cite{ciarletta2016residual}. Thus, it would be possible to derive the dispersion curves governing the propagation of  time-harmonic spherical waves of small amplitude  as a function of the residual stress components. This theoretical prediction may feed  a nonlinear inverse analysis for determining the pre--stress distribution using elastic waves, e.g. by ultrasound elastography \cite{Man1987,li2017guided}.

Furthermore, our results prove useful insights for designing  mechanical meta-materials  with adaptive morphology.  Indeed, it would be possible to fabricate soft spheres in which the magnitude of the pre-stresses can be controlled by external stimuli, such as voltage in dielectic elastomers  \cite{brochu2010advances} or solvent concentration in soft gels \cite{tokarev2009stimuli}. Digital fabrication techniques  offer a low cost alternative for printing materials with a targeted distribution of residual stresses \cite{zurlo2017printing}.  Thus, morphable spheres can be obtained by modulating the residual stresses around the  critical value of marginal stability. Dealing with pre--stressed neo-Hookean materials, this work is particularly relevant for controlling the transient wrinkles that form and then vanish during the drying and swelling of hydrogels \cite{lucantonio2013transient,bertrand2016dynamics}.  Other applications range from adaptive drag reduction \cite{terwagne2014smart} to the pattern fabrication on spherical surfaces \cite{stoop2015curvature,brojan2015wrinkling}.

\section{Acknowledgments}

This work is funded by AIRC MFAG grant 17412. We are thankful to Simone Pezzuto and Matteo Taffetani for useful discussions on numerical issues. D.R. gratefully acknowledges funding provided by \emph{INdAM--GNFM} (National Group of Mathematical Physics) through the program \emph{Progetto Giovani 2017}.

\bibliographystyle{SageV}
\bibliography{refs-main}
\end{document}